\documentclass[12pt]{article}
\usepackage{graphicx} % Required for inserting images
\usepackage{booktabs}
\usepackage[square, numbers]{natbib}
\usepackage{amsmath}
\usepackage{xr}
\usepackage{xcite}
\usepackage[margin=25.4truemm]{geometry}
\usepackage{lineno}
\usepackage{authblk}
\usepackage{hyperref}
\usepackage{lineno}
\usepackage{float}
\hypersetup{colorlinks, allcolors=black}

\date{}

\title{Prestige bias drives the viral spread of content reposted by influencers in online communities}
\author[1]{Takuro Niitsuma*}
\author[2]{Mitsuo Yoshida}
\author[1]{Hideaki Tamori}
\author[3]{Yo Nakawake*}
\affil[1]{Media R\&D Center, The Asahi Shimbun Company, 5-3-2 Tsukiji, Chuo-ku, Tokyo, Japan}
\affil[2]{Institute of Business Sciences, University of Tsukuba, 3-29-1 Otsuka, Bunkyo-ku, Tokyo, Japan}
\affil[3]{Graduate School of Advanced Science and Technology, Japan Advanced Institute of Science and Technology, 1-1 Asahidai, Nomi, Ishikawa, 923-1211, Japan}



\begin{document}
	
	\maketitle
	\par
	
	*Corresponding authors: T.N. (niitsuma-t@asahi.com) and Y.N. (nakawake@jaist.ac.jp).
	%TC:ignore
	\abstract{
		Cultural evolution theory suggests that prestige bias—whereby individuals preferentially learn from prestigious figures—has played a key role in human ecological success. However, its impact within online environments remains unclear, particularly with respect to whether reposts by prestigious individuals amplify diffusion more effectively than reposts by noninfluential users. We analyzed over 55 million posts and 520 million reposts on Twitter (currently X) to examine whether users with high influence scores (hg indices) more effectively amplified the reach of others' content. Our findings indicate that posts shared by influencers are more likely to be further shared than those shared by non-influencers. This effect persisted over time, especially in viral posts. Moreover, a small group of highly influential users accounted for approximately half of the information flow within repost cascades. These findings demonstrate a prestige bias in information diffusion within the digital society, suggesting that cognitive biases shape content spread through reposting.
	}
	%TC:endignore
	\newpage
	\section{Introduction}
	In the digital age, social media influencers have emerged as authoritative figures with significantly greater information diffusion power than ordinary users do~\cite{Pei2014,Kempe2003}. The impact of influencers extends beyond the online world, as evidenced by the emergence of influencer-based viral marketing strategies~\cite{Yesiloglu2020,Zietek2016,Watts2007} and legal disputes concerning reposts by an influencer on Twitter (currently X)~\cite{Murakami2021}.
	Influencers are believed to significantly impact the distribution of information in online communities because of their voice and follower count.
	In recent years, various approaches have been used to address problems related to online communication, such as the spread of false information or homogeneity of received information (so-called echo chambers and filter bubbles).
	However, focusing on influences remains crucial in these contexts.
	Research has shown that influencers, particularly experts in specific areas, can contribute to correcting false information in online communities~\cite{Lim2022}.
	Accordingly, a continued focus on influencers remains crucial in this context.
	
	Previous research examining influencers' role in information diffusion has focused on their capacity as \textit{source spreaders}; i.e., their ability to generate and spread original content~\cite{Pei2014,Morone2015,Kempe2003,Tsugawa2023}.
	However, a key feature of modern social networks comprises repost or share functions, which allow users to reproduce others' posts; such functionality is a major contributor to information diffusion. \citet{Acerbi2022} noted that sharing content online fundamentally differs from oral retelling, as it does not require memorization or reproduction of the content. This facilitates and potentially accelerates the spread of information in online environments, contributing to viral-like diffusion patterns.
	Some researchers have examined influencers as \textit{brokers} of information, analyzing their role in facilitating the spread of content created by others~\cite{Burt2000,TheoAraujo2017,Tsugawa2023}. Despite these research advances, significant gaps remain.
	Among these is the question of whether influencers who excel at spreading self-generated content can also amplify other-generated content through reposts.
    That is, are recipients more likely to share a post simply because an influencer has reposted it?
	Thus, the present study aims to investigate whether the diffusion capabilities of influencers extend beyond their own content, potentially affecting the propagation of information posted by others. Specifically, this research aims to explore how these influencers impact the preferences and engagement behaviors of users who receive shared content. %%% 追加
	
	% 受け手にとってインフルエンサーの方がrepostしやすいということは、そこに認知バイアスあると考えられると
	% 文化進化には、context-dependent biasと呼ばれており、心理学や人類学を中心として研究されてきた
	If recipients are indeed more likely to share content from influencers, this would suggest the presence of cognitive biases. In evolutionary anthropology and cultural psychology, researchers have long studied context-dependent biases in information acquisition, where individuals selectively adopt information on the basis of social context rather than content alone.
	Among these, prestige bias is the tendency to learn from socially successful or prestigious individuals~\cite{Henrich2001,Jimenez2019,Henrich2003}; it is considered an adaptive trait from a historical perspective.
	This concept aligns with established theories such as the theory of communication (e.g., two-step flow\cite{Katz1955} and diffusion of innovations\cite{Rogers2003}), which emphasize the role of opinion leaders and early adopters in information spread.
	Recent studies on social media diffusion patterns~\cite{Goel2012} have indicated that most sharing occurs within close proximity to the original poster, highlighting the importance of early-stage diffusion and influential users. Furthermore, research has shown influencers' potential to correct false information~\cite{Lim2022}.
	Building on these theories and recent findings, this study focuses on the early stages of information diffusion.
	We aim to track and analyze how influence changes over time, examining the effect of prestige bias in the digital contexts characterizing our information-rich society.
	
	When applied to social media, the concept of prestige bias raises important questions: do influencers function as prestigious models whose reposts are more likely to be further reposted than noninfluencers? Clearly, influencers' original content appeals to their audience of active followers; however, it remains uncertain whether influencers maintain a similar influence when merely reposting others' information. Understanding whether there is a cognitive tendency to prefer any information associated with prestigious individuals, whether original or reposted, would provide useful insights into how information recipients are influenced.
	Our research thus aims to investigate whether having a proven ability to spread original content (our definition of "influencer") translates into an enhanced ability to spread others' content through reposts. Importantly, this study does not focus on the number of followers an influencer has; rather, it examines individual cognition and behavior. Specifically, it examines whether individuals are more likely to spread information they receive from influencers than from noninfluencers.
	Our investigation of the early stages of information diffusion allows us to explore the nuanced dynamics of information spread in social media, building insights into the cognitive processes underlying user behavior (content sharing) beyond simple metrics.
	
	\begin{figure}[ht]
		\centering
		\includegraphics[width=\textwidth]{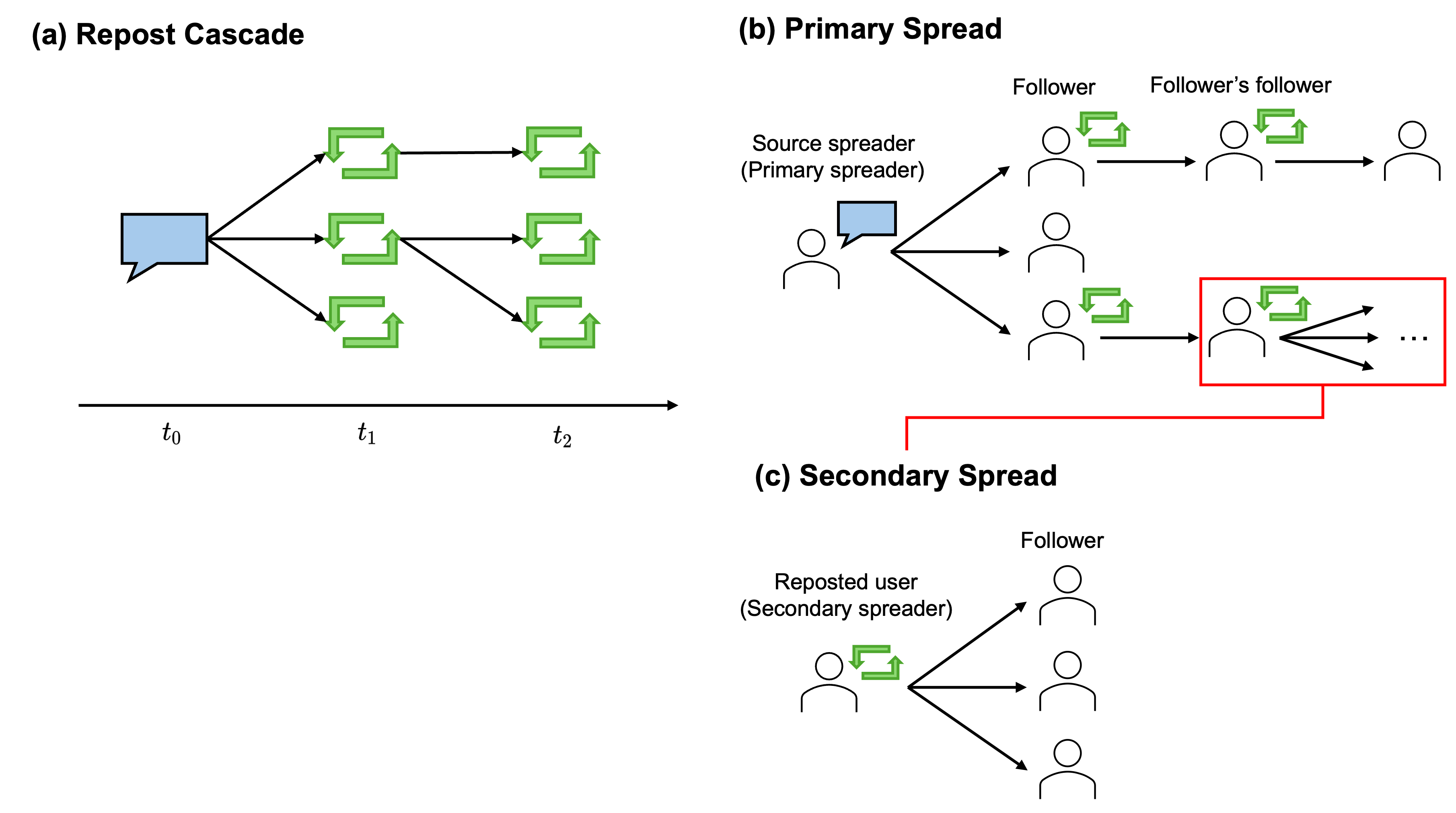}
		\caption{(a) Illustration of a repost cascade. The blue speech bubble represents an original post, whereas the green repost icons represent subsequent reposts. The arrows indicate the direction of information flow. (b) Illustration of primary spread, where an original post (blue speech bubble) is directly shared with multiple users. (c) Illustration of secondary spread, where a user reposts (green icon) content from another user, which is then further shared.}
		\label{fig:repost_cascade}
	\end{figure}
	
	To test our hypothesis about the existence of prestige bias in online communities, we need to precisely track how information flows to subsequent recipients via reposts; this requires the introduction of several key concepts.
	First, we introduce the concept of a \textit{repost cascade}, which builds upon the established notion of information cascades~\cite{Bikhchandani1992,Watts2002,Kempe2003,Goel2012,Vosoughi2018}. Figure \ref{fig:repost_cascade} illustrates this concept and shows two key types of information spread that we distinguish in our study. This framework allows us to identify who received information from whom.
	As shown in Figure \ref{fig:repost_cascade}(a), a repost cascade represents the chain of reposts originating from an original post, capturing the cascading structure of information diffusion on social networks. The blue speech icon (speech bubble) represents an original post, whereas the green repost icons represent subsequent reposts, with arrows indicating the direction of information flow over time.
	Next, we distinguish between two types of information spread: \textit{primary spread}, which is the diffusion of a user's original post, and \textit{secondary spread}, which is the diffusion of a user's repost.
	Figure \ref{fig:repost_cascade}(b) depicts the primary spread, where an original post (blue speech bubble icon) from a source spreader is directly shared with multiple users, including followers and followers' followers. This process represents the propagation of original content through the network.
	The secondary spread, shown in Figure \ref{fig:repost_cascade}(c), represents the phenomenon whereby content is reposted by a user (indicated by the green repost icon) and then further spread among their followers. This secondary process captures how information continues to spread beyond its initial audience through the actions of intermediary users.
	This distinction is crucial for our analysis, as it allows us to differentiate between a user's ability to spread their own content (primary spread) and their capacity to amplify others' messages (secondary spread). By focusing on these two types of spread, we can investigate whether accounts with powerful primary spread capabilities (influencers, which are quantified by the hg index\cite{Alonso2010} in our study) also have powerful secondary spread capabilities.
	
	\begin{figure}[htp]
		\centering
		\includegraphics[width=160 mm]{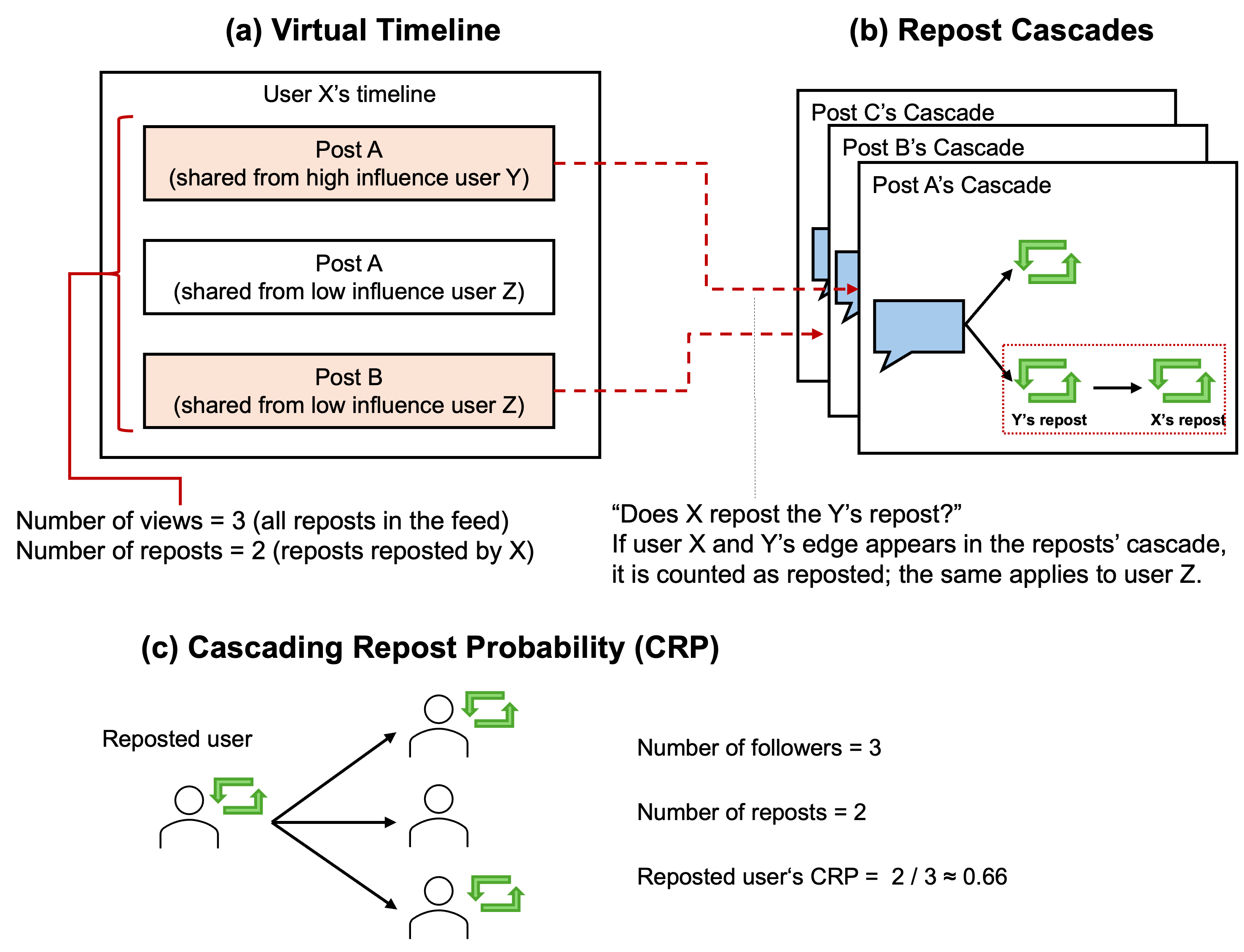}
		\caption{(a) Virtual timeline: A representation of posts shared by users with different influence levels. (b) Repost cascades: Illustration of how reposts are tracked in cascades. If the edge linking users X and Y appears in the post's cascade, it is counted as having been reposted. (c) Cascading repost probability (CRP): An example calculation of the CRP for a reposted user with three followers, where two of them repost, resulting in a CRP of $2/3 \approx 0.66$.}
		\label{fig:virtual_timeline_crp}
	\end{figure}
	Furthermore, to track who views and engages with reposts from specific users, we construct a virtual timeline from the sampled reposts and follower‒followee relationships, as illustrated in Figure \ref{fig:virtual_timeline_crp}(a). This approach allows us to simulate followers’ exposure to reposts and observe their subsequent engagement.
	Finally, via this approach, we introduce the concept of cascading repost probability (CRP) to measure the efficiency of information spread (Figure \ref{fig:virtual_timeline_crp}(b,c)).
	We consider a repost on the virtual timeline as having been further shared if it appears in a repost cascade, allowing us to calculate the probability of a user's secondary spread continuing (Figure \ref{fig:virtual_timeline_crp}(b)).
	The CRP quantifies the likelihood of a repost being further shared, enabling us to evaluate the efficiency of repost diffusion among users with varying levels of influence; the CRP thus measures a user's influence in the context of secondary spread. (Details are shown in Figure \ref{fig:virtual_timeline_crp}(c).)
	
	We calculate the CRP by aggregating the view and repost counts across all the users' timelines.
	By comparing the CRPs of influencers and noninfluencers, we can assess whether prestige bias manifests in online environments. If influencers consistently demonstrate higher CRP values, it suggests that their status enhances their ability to spread information, even when that information originates from others.
	
	These concepts (i.e., repost cascade, secondary spread, virtual timeline, and CRP) provide us with a framework to quantify and analyze users' influence in terms of secondary information spread. Our methodology focuses on the dynamics of secondary spread and how it relates to users' level of influence, particularly when comparing influencers and noninfluencers.
	On the basis of this framework, we hypothesize that influencers have higher CRP values than noninfluencers; these values remain consistently higher over time, although they are especially high in the early stages after the original post has been posted.
	Moreover, we also analyzed this hypothesis with a mixed-effects logistic regression that controlled for post-level and user-level heterogeneity as random effects.
	If supported, this would provide evidence for the effect of prestige bias in online information diffusion.
	Notably, while the CRP represents the efficiency of reposts in the secondary spread, it does not capture the actual scale of distribution and indirect influence. Therefore, to further support our hypothesis, we also identified the proportion of the actual distribution of reposts in secondary spreads attributed to influencers and analyzed the impact on max depth and structural virality of a cascade from first-reposted users.
	This comprehensive approach allows us to examine the efficiency of influencers’ information spread and its overall impact on information diffusion in online communities.
	
	We thus aim to provide insight into how user status and influence shape the dynamics of information spread in digital environments.
	Our study thus contributes to the broader understanding of information diffusion in online social networks, the role of influencers in this process, and the potential presence and impact of prestige bias in digital contexts. The results of this investigation have implications for our understanding of online information dynamics, influencer marketing strategies, and the design of social media platforms.
	
	\section{Results}
	
	\subsection{Data Collection}
	Following established methodologies in previous studies~\cite{Goel2012,Vosoughi2018}, we constructed repost cascades on the basis of one month of Japanese-language posts sampled from Twitter (currently X) and their associated follower‒followee relationships. The dataset comprises 55,882,528 source posts, 520,048,995 reposts, and 14,910,772 unique users. For detailed information on the data collection process and the rationale behind choosing Japanese-language posts, please refer to the subsection \ref{hg index} in the Methods.
	
	\subsection{User Influence Distribution and Source Post Popularity}
	\label{subsec:crp}
	
	To categorize users by their influence as source spreaders, we employed the hg index, which quantifies a user's ability to consistently generate and spread original content. This metric extends the h index~\cite{Hirsch2005}, which is commonly used to measure scientific productivity, by incorporating additional factors. Thus, it offers a more comprehensive measure of a user's primary spread capability (see details in Methods). On the basis of the distribution of the hg index scores, we classified users into six influence categories through quantile binning: very high (top 1\%), high (top 1--5\%), upper-mid (top 5--10\%), mid (top 10--30\%), lower-mid (top 30--50\%), and low (bottom 50\%). Each category excludes its upper threshold.
	Table \ref{tab:user_influence} summarizes these user categories, showing the number of users, their average hg index, and the popularity of their original posts. As indicated, users in the very high category exhibit the highest average repost counts as well as the highest maximum number of reposts.
	
	\begin{table}[ht]
		\centering
		\caption{Influence scores as source spreaders and popularity of original posts by user category. User influence, No. of users, Avg. hg index, Avg. reposts (mean, average number of reposts per post for each user category). , Freq. of reposts (the maximum number of reposts received by a single post within each category). \\ }
		\label{tab:user_influence}
		\begin{tabular}{lrrrr}
			\hline
			User influence & No. of users & Avg. hg index & Avg. reposts & Freq. of reposts  \\
			(category) & (count) & (mean) & (mean) & (maximum) \\
			\hline
			very high & 160,719 & 19.89 & 24.48 & 140,856 \\
			high & 545,992 & 5.69 & 4.39 & 63,496 \\
			upper-mid & 597,749 & 2.87 & 2.58 & 28,301 \\
			mid & 897,878 & 1.80 & 2.00 & 23,518 \\
			lower-mid & 3,416,185 & 1.00 & 1.48 & 16,407 \\
			low & 10,725,364 & 0.00 & 0.00 & 0 \\
			\hline
		\end{tabular}
	\end{table}
	
	\subsection{Analyzing Prestige Bias in Secondary Spread}
	We next explored how user influence impacts the secondary spread of information---that is, how content diffuses after being reposted by users other than the original poster. We measure this chain of secondary spread via the cascading repost probability (CRP). First, we construct a virtual timeline for each user, tracking when reposts appear. Then, to determine whether a repost occurred, we inspect repost cascades: if user Y's repost appears in user X's virtual timeline, we check whether X subsequently shares the content. If so, we count it as having spread by Y. Finally, we aggregate the repost counts by users' influences. Figure \ref{fig:virtual_timeline_crp}(b) illustrates this process. Because we focus on initial diffusion, we aggregate the results over the first 6 h after a post appears.
	
	\begin{figure}[htbp]
		\centering
		\includegraphics[width=0.75\textwidth]{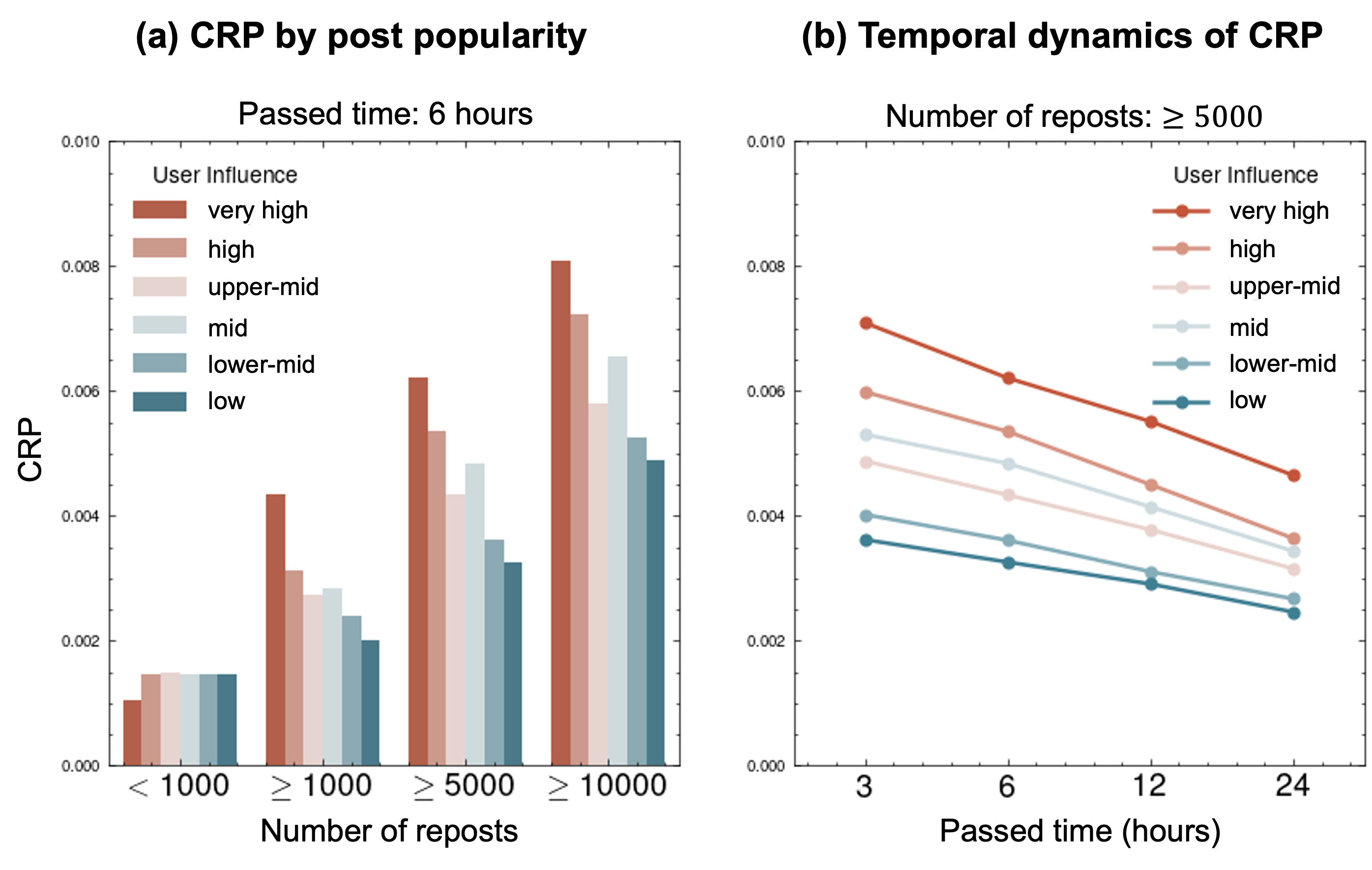}
		\caption{Analysis of the cascading repost probability (CRP). (a) CRPs by post popularity for different user influence categories 6 h after posting. (b) Temporal dynamics of CRPs over 24 h for posts with $\geq$ 5000 reposts. The x-axis indicates hours since the original post was published. In both panels, very high-influence users consistently show higher CRP values across all popularity thresholds except $<1000$; this advantage is maintained over time.}
		\label{fig:crp_analysis}
	\end{figure}
	%%%
	
	Figure \ref{fig:crp_analysis}(a) shows the CRP within the first 6 h after posting, categorized by the minimum number of reposts ($<$ 1000, $\geq$ 1000, $\geq$ 5000, $\geq$ 10000) and user influence categories. The results demonstrate that users with very high influence consistently exhibit higher CRP values for posts with a high number of reposts ($\geq$ 1000), indicating a greater ability to propagate information, even when they are not the original source. Notably, the difference in the CRP between influence categories becomes more pronounced as the minimum number of reposts increases, suggesting that the impact of user influence is particularly strong for highly popular content.
	
	To examine the temporal dynamics of this effect, we analyzed the CRP over a 24-h period after the original post time for cascades with $\geq$ 5000 reposts, as shown in Figure \ref{fig:crp_analysis}(b). The x-axis in Figure 3(b) represents hours elapsed since the original post. The graph displays the CRP for different influence categories over 24 h. Although the CRP generally decreases over time for all categories, users with very high influence maintain substantially higher CRP throughout the observation period, further emphasizing their sustained impact on information diffusion.
	
	These findings collectively demonstrate that user prestige enhances the perceived value or interest of shared content, thereby increasing its likelihood of further diffusion.
	Our analysis shows that prestige bias plays a pivotal role in online social networks, which is consistent with our initial hypothesis. In particular, users with greater influence, as measured by the hg index, exhibit a greater capacity to propagate information, even if it originates from others.
	
	\begin{figure}[htbp]
		\centering
		\includegraphics[width=0.8\textwidth]{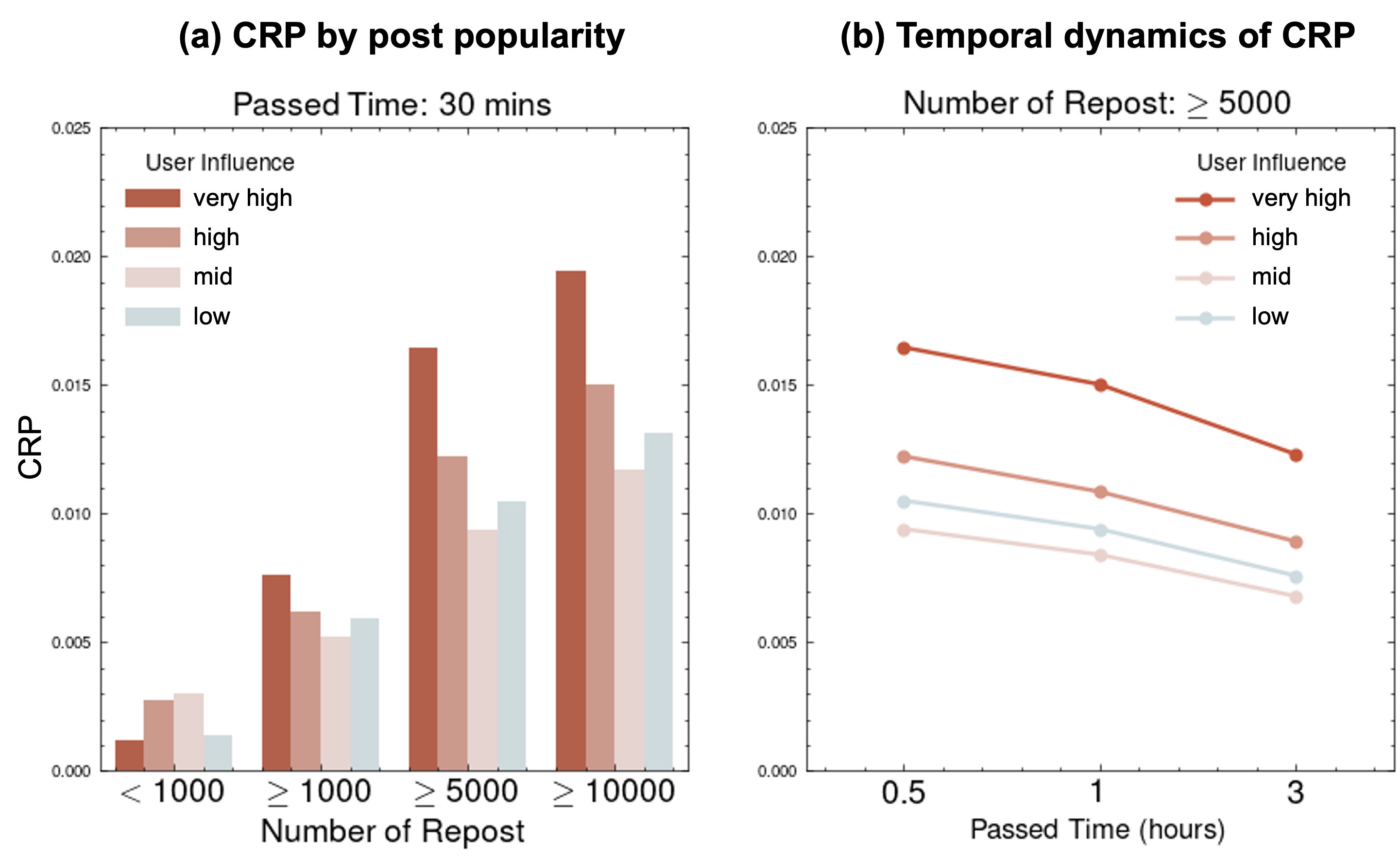}
		\caption{Cascading repost probability (CRP) analysis using English-language posts collected on January~1, 2015. The left panel (a) shows the results restricted to 30\,min, and the right panel (b) compares 30\,min, 1\,h, and 3\,h.}
		\label{fig:crp_analysis_en}
	\end{figure}
	
	Finally, to verify the robustness of our results, we examined a dataset of English-language posts collected on January 1, 2015 (see Supplementary Material S1). % \ref{sup-english-data}
	We used a sample of one million users from December 2014, provided by \citep{Yamaguchi2015}\cite{Yoshida2015}, and collected posts from X (formally Twitter)’s stream over 24 h on January 1. Because this time window is shorter than in our main analysis, we focused on shorter intervals.
	In Figure \ref{fig:crp_analysis_en}, panel (a) shows the CRP during the first 30\,min, and panel (b) compares the results after 30\,min, 1\,h, and 3\,h. Despite this limited period, a consistent pattern was found: very high-influence users still exhibit high CRP.
    Interestingly, in this dataset, the \textit{low} category outperforms \textit{mid}.
    This is possibly due to misclassifications, as the hg index is calculated from a single day’s data (some influential users may not have posted on that day). Moreover, this English-language dataset captures only one day of user behavior, so it may not fully capture longer-term diffusion patterns.
    Despite these minor differences, the analysis overall replicates our main findings in the Japanese dataset: users with greater influence tend to be more effective at propagating content, even under different linguistic and temporal conditions.
	
	\subsection{Analysis via Mixed-Effects Logistic Regression}
	\label{seubsec:melr}
	In the analysis presented in subsection \ref{subsec:crp}, we did not consider the possibility that influencers select potentially attractive posts that could be reposted by other users. Furthermore, we did not consider the effects of specific posts or users. To exclude these confounding factors, we applied mixed-effects logistic regression accounting for both user-level and post-level heterogeneity.
	Because the dataset is fairly large, a sampling procedure was used to retain approximately 1\% of all positive (reposted) cases along with a twofold number of negative cases. The complete details of the models, dataset, and sampling approach are provided in the Methods section.
	
	In this model, the dependent variable is a binary indicator of whether the user who received a repost subsequently reposts it. The model includes the original reposting user’s influence category (\texttt{sender\_influence}), the repost’s hour of the day (\texttt{repost\_hour}), and log-transformed and standardized follower counts for both the user who reposted a post (\texttt{sender\_followers\_count}) and the user who received the repost (\texttt{user\_followers\_count}). Random intercepts are introduced for each original tweet (\texttt{source\_tweet\_id}) and the topic of the user profiles (\texttt{user\_topic}), thereby controlling for differences in content appeal and individual interests.
	
	\begin{table}[h]
		\centering
		\caption{Fixed-effects estimates from the mixed-effects logistic regression (sender\_influence baseline = \textit{low}).}
		\label{tab:regresults}
		\begin{tabular}{lrrrr}
			\toprule
			\textbf{Predictor} & \textbf{$\beta$ value} & \textbf{Std. Error} & \textbf{$z$ value} & \textbf{$p$ value} \\
			\midrule
			(Intercept)                           & -0.927 & 0.137 & -6.777   & $<$0.001 \\
			\texttt{sender\_influence1} (lower-mid) & 0.148 & 0.006 & 23.085  & $<$0.001 \\
			\texttt{sender\_influence2} (mid)     & 0.246 & 0.007 & 37.489  & $<$0.001 \\
			\texttt{sender\_influence3} (upper-mid)& 0.287 & 0.006 & 45.094   & $<$0.001 \\
			\texttt{sender\_influence4} (high)    &  0.387 & 0.006 &  66.123   & $<$0.001  \\
			\texttt{sender\_influence5} (very high)&  0.599 & 0.006 & 99.480   & $<$0.001 \\
			\texttt{repost\_hour}: morning           & -0.027 & 0.003 & -8.227   & $<$0.001 \\
			\texttt{repost\_hour}: noon              &  0.034 & 0.003 &  9.813   & $<$0.001 \\
			\texttt{repost\_hour}: night             &  0.031 & 0.003 & 10.010   & $<$0.001 \\
			\texttt{repost\_hour}: midnight         & -0.018 & 0.004 & -4.626   & $<$0.001 \\
			\texttt{sender\_followers\_count}        & -0.165 & 0.001 & -114.126 & $<$0.001 \\
			\texttt{user\_followers\_count}          & -0.568 & 0.001 & -461.838 & $<$0.001 \\
			\bottomrule
		\end{tabular}
	\end{table}
	
	Table~\ref{tab:regresults} presents the fixed-effects estimates for the model, with low influence serving as the baseline. The influence categories all have positive coefficients relative to this baseline, and the very high category has a substantial positive coefficient. Although these estimates cannot be used to infer absolute repost probabilities because of our sampling ratio, the relative differences between categories are meaningful. The effect of \textit{very high} ($\beta = 0.599$) suggests that top-tier influencers can markedly amplify the likelihood of secondary spreading, which is consistent with a prestige bias in online diffusion.
	The other predictors show comparatively modest but significant effects. Variations by hour of day point to diurnal patterns in user engagement, with slight increases in reposting at night and noon. Both \texttt{sender\_followers\_count} and \texttt{user\_followers\_count} have negative coefficients, indicating that once the influence category and other factors are considered, having a larger follower base does not necessarily promote further reposting of others’ content.

    Furthermore, our analysis indicates that the random intercept for \texttt{source\_tweet\_id} exhibits a large variance, suggesting that content-specific factors play an important role in online diffusion.
    However, our main hypothesis focuses on the effect of user influence rather than the content-specific aspects of the posts.
    Therefore, we do not present an analysis of content-specific factors in the main text. Nevertheless, we present further details of a topic-specific analysis in the Supplementary Material S3 for reference. % \ref{sup-content-analysis}
	
	\subsection{Quantifying the Impact of Influencers on Secondary Spread}
\begin{figure}[htbp]
		\centering
		\includegraphics[width=140 mm]{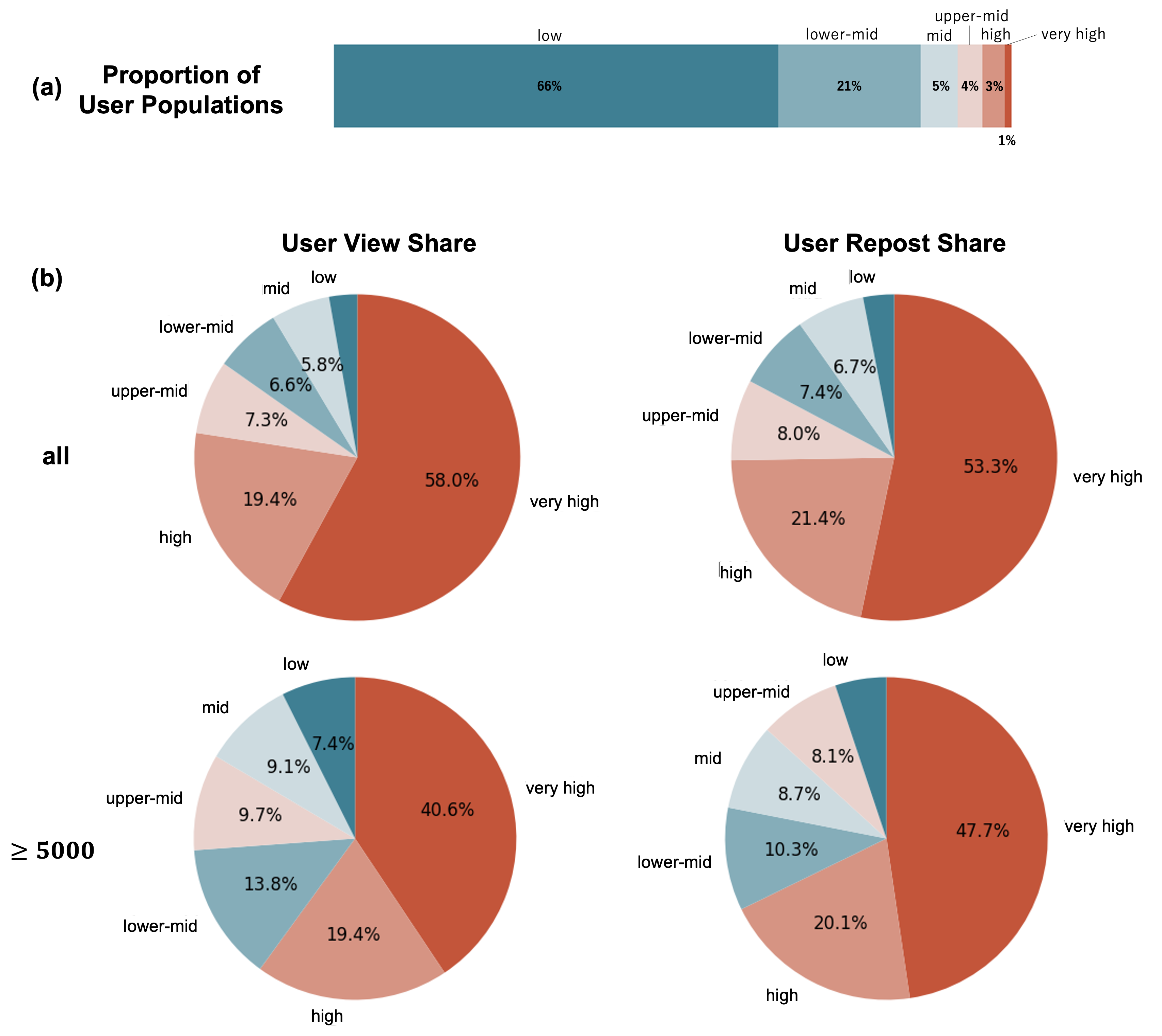}
		\caption{Distribution of secondary spread across influence categories. (a) Proportion of users according to influence levels (for simplicity, percentages are rounded to integers). (b) User view share (proportion of views for each influence category). and user repost share (proportion of reposts that can be traced back to each influence category). The top row shows data for all posts, whereas the bottom row focuses on posts with $\geq$ 5000 reposts. Notably, very high-influence users (the top 1\% of the user population) consistently account for approximately half of the views and repost in the secondary spread, demonstrating their substantial impact on information distribution.}
		\label{fig:user_share}
	\end{figure}

	Although the CRP is a useful indicator of information propagation efficiency, it does not fully capture the scale of that propagation, which can be measured by the actual quantity of reposts. To address this limitation and provide a more comprehensive analysis of influencers' role in shaping information flow through secondary spread, we investigated the proportion of views and reposts in each influencer category.
	In our analysis, we calculate the number of times users potentially see shared posts in their simulated timelines, which we refer to as \textit{views}; these views represent instances where a user would encounter a repost in their timeline (Figure \ref{fig:virtual_timeline_crp}(a)).
	When analyzing the secondary spread in the virtual timeline, it is crucial to distinguish between these views and reposts, as views only indicate exposure to content, whereas reposts represent active propagation of the content to other users.
	By tracking views, we can determine how often reposts (secondary spread) by users with different influence levels are potentially seen in other users' timelines. This allows us to quantify not only how many times content is reposted but also its potential visibility across the repost cascade. Therefore, our approach provides insights into both the spread efficiency (measured by the CRP) and the potential reach of information shared by different user categories.
	Figure \ref{fig:user_share} illustrates the proportion of users according to influence levels as well as the distribution of views and reposts in the secondary spread according to influence categories.
	
	The results reveal that users with very high influence, despite comprising only 1\% of the user base (Figure \ref{fig:user_share}(a)), account for 58.0\% of views and 53.3\% of reposts in secondary spread across all posts (Figure \ref{fig:user_share}(b) above). This disproportionate influence becomes somewhat less pronounced when the number of reposts is large ($\geq$ 5000), where very high-influence users are responsible for 40.6\% of views and 47.7\% of reposts (Figure \ref{fig:user_share}(b) bottom).
	Notably, for highly popular posts ($\geq$ 5000 reposts), an interesting shift occurs in the behavior of very high-influence users. Although their overall share of both views and reposts decreases compared with all posts, their share of reposts (47.7\%) now exceeds their share of views (40.6\%). This contrasts with the pattern observed for all posts, where their share of views (58.0\%) is greater than their share of reposts (53.3\%). This reversal is particularly significant, given that reposts can only occur after a user views the content.
	This finding not only aligns with the high CRP observed for this category in Figure \ref{fig:crp_analysis}(b) but also indirectly demonstrates that when influencers share content, it generally receives high levels of engagement from their followers, thereby contributing to its popularity.
	In this way, the role of very high-influence users shifts with respect to viral content. Instead of merely exposing content to their large follower base, they effectively amplify the content through their own sharing (by encouraging reposting).
	This amplifying effect of influencers plays a crucial role in accelerating and expanding the propagation of popular content through secondary spread.
	
	\subsection{Influence of the First-Reposted Users on Viral Diffusion}
	\label{subsec:results_influence}
	Thus far, we have focused only on the direct followers of influencers; however, influencers can further affect nondirect followers (followers' followers) through a repost cascade. Here, we further analyze the depth and complexity of the repost cascade.
	
\begin{figure}[htbp]
		\centering
		\includegraphics[width=0.8\textwidth]{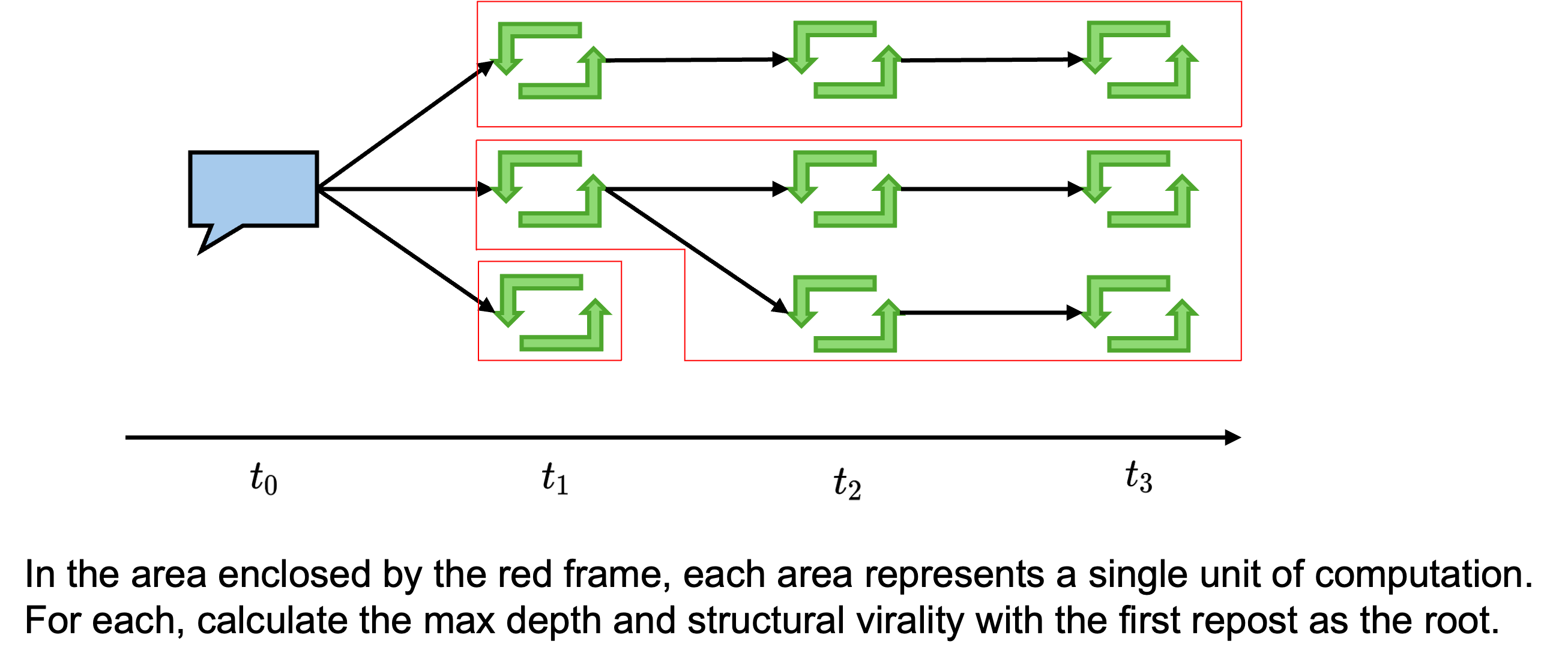}
		\caption{Conceptual repost subtree rooted at the first-reposted users. Rectangular regions represent progressively deeper layers of sharing.}
		\label{fig:conceptual_subtree}
	\end{figure}

	An analysis of repost cascades for posts that attracted more than 5,000 reposts revealed a strong relationship between the influence level of the first user who shared a post and its diffusion depth and complexity. Figure~\ref{fig:conceptual_subtree} illustrates the conceptual structure of a repost subtree, with each rectangular layer representing a stage in the diffusion process. When high- or very high-influence users make the initial repost, the content typically propagates through more tiers of sharing, resulting in greater structural virality and an increased maximum depth.
	
\begin{figure}[htbp]
		\centering
		\includegraphics[width=\textwidth]{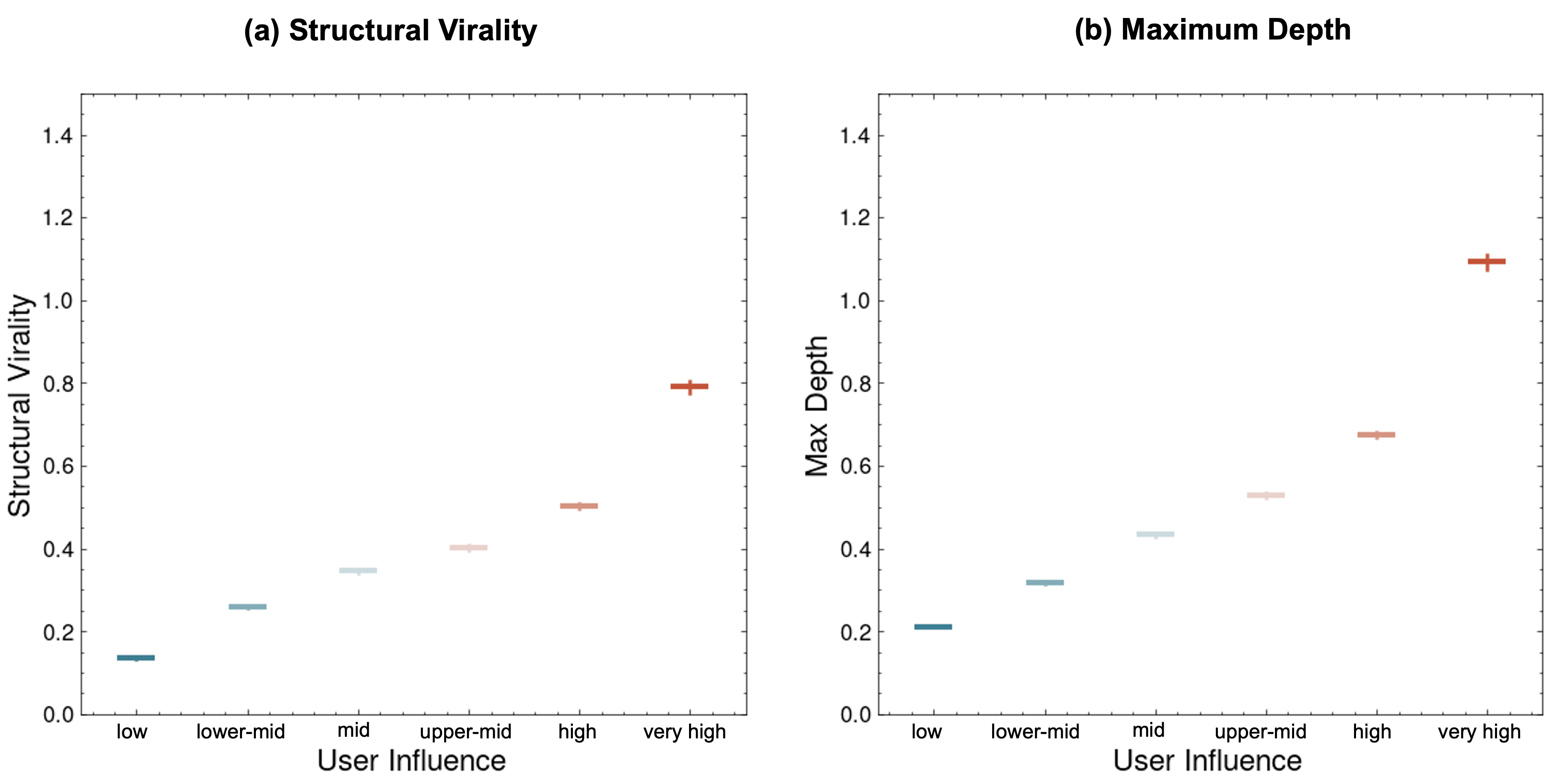}
		\caption{(Left) structural virality and (right) maximum depth across different user influence levels. The error bars represent confidence intervals for each influence category.}
		\label{fig:virality_depth}
	\end{figure}

	Structural virality was measured by calculating the average path length among all pairs of users within each repost cascade. This analysis provided a sense of how extensively and how many steps the content traveled. The maximum depth was determined by measuring the longest path from the first-reposted users to the most distant individual in the repost chain. As shown in Figure~\ref{fig:virality_depth}, both metrics tended to increase with increasing user influence. In particular, the \textit{very high} category presented the largest values, indicating that these individuals—often characterized as influencers—did more than simply broadcast posts to a large audience. They appeared to promote multistep diffusion cascades, suggesting that followers who received the post from a highly influential source were more inclined to share it further, extending its overall reach and complexity.
	These findings underscore the importance of accounting for user influence in viral diffusion.
	
	\section{Discussion}
	In this study, we introduced new concepts (secondary spread and CRP) to test the hypothesis that influencers (users with high hg index scores) more effectively propagate information when resharing others' content than noninfluencers do; this analysis was conducted using one month of Japanese-language post data.
	
	Our analysis revealed that influencers consistently demonstrate significantly higher CRP for popular posts ($\geq$ 1,000 reposts) in secondary spreads than noninfluencers. The CRP, indicating influencers' impact, remained high over time, showing a robust, sustained influence on information diffusion. Moreover, influencers were found to substantially impact the actual distribution of reposts in secondary spread. Over half of the views and reposts in secondary spreads could be attributed to the top 1\% of users, with this trend being even more pronounced for highly popular posts. One might argue that the CRP value is inherently low in absolute terms (e.g., 0.01). However, this is likely because we count all reposts appearing in the virtual timeline as views. If we could evaluate only the posts that users actually viewed, the probability might be slightly higher. Nevertheless, as suggested by previous studies ~\citep{Goel2012,Watts2002}, consecutive reposts occurring in sequence remain rare phenomena. Even if the absolute term of the CRP is low, its relatively high value in the early stages can still be considered a crucial metric if it influences the subsequent size of the cascade for that post.
	Note that these effects are not due to more frequent reposting by highly influential users. The analysis in Supplementary Material S4 demonstrates that very high-influence users do not repost more often than users in other influence categories. Instead, the impact comes from the wider reach of influential users' reposts and the greater likelihood of those reposts being shared further (higher CRP). % \ref{sup-reposting-behavior}
	Moreover, these results were replicated on the one-day post dataset of the English language.
	
	A possible concern is that relying solely on a strictly chronological sequence may overlook the impact of engagement and advertising.
	We restricted our analysis to the first 30 min after a post’s publication to reduce the influence of algorithmic recommendations.
	Many early-stage posts are unlikely to have significant differences in engagement, which may help mitigate the influence of algorithms.
	The results (Supplementary Material S5) remained consistent with those reported here, reinforcing the reliability of our virtual timeline approach. % \ref{sup-sensitivity-30mins}
	
	Additionally, the results of the mixed effects logistic regression controlling for post-level and user-level heterogeneity were consistent with the results of the CRPs in subsection \ref{subsec:crp}.
	
	Our findings provide consistent evidence for the effect of prestige bias in online social networks. Users with greater influence, as measured by their hg index, consistently demonstrate a greater ability to propagate information, even when that information is not their own original content. This effect persists throughout the important early stages of information diffusion.
	These results align with the theoretical framework of prestige bias proposed in anthropology and psychology~\cite{Henrich2001,Jimenez2019}, suggesting that this cognitive tendency extends to digital environments. Influential users’ consistently higher CRP values indicate that their status enhances their ability to spread information, supporting our initial hypothesis.
	The novelty of this research lies in its integration of the roles of influencers as source spreaders (originators of content) and brokers (information intermediaries), which have been the focus of previous influencer studies. By introducing the concepts of secondary spread and the CRP, we reveal how influencers effectively fulfill both roles, providing a new perspective that bridges existing research areas. Furthermore, this study empirically demonstrates the effect of prestige bias in online communities. We have shown that information reposted by influencers tends to spread more widely and persistently.
	
	Interestingly, influencers did not demonstrate much influence in relation to unpopular posts. This finding suggests that prestige bias may depend on both influencer and content characteristics. Traditional theories assume that prestige bias uniformly affects all kinds of information; our results suggest that this may not be the case in online communities.
	This implication is also consistent with the results that the variance of \texttt{source\_tweet\_id} is greater than the \texttt{user\_influence} coefficients of \textit{very high} in the mixed-effects logistic regression analysis, which is detailed in Supplementary Material S3. % \ref{sup-content-analysis}
	In this context, the research by \citet{Acerbi2018} provides important insights. Their experiments revealed that in selecting quotations, the content was more important than their attribution to famous individuals. This result suggests that in online environments, the quality of content may be more pertinent in evaluating information than the status of the sender.
	However, our results also suggest that the content of the shared information may become important when the content has a strong appeal. That is, prestige bias may play a stronger role when the quality of content is equivalent.
	Moreover, although previous research, such as that by \citet{Brand2021}, has shown that source expertise influences the perceived reliability of information on social media, our findings significantly extend this understanding.
	
	Our analysis of structural virality~\cite{Goel2016} in widely diffused posts reveals that influencers' impact actively shapes information diffusion patterns.
	Posts reposted by influencers not only reach a wider audience but also exhibit greater structural virality, indicating more complex and extensive diffusion. This, coupled with the high CRP observed among influencers for potentially viral content, suggests a dual role for influencers: they act as both broadcasters, directly reaching large audiences, and amplifiers, significantly increasing the likelihood of further sharing by their followers.
	By enhancing the \textit{shareability} of their posts (and reposts), influencers create a ripple effect extending beyond their direct connections. Identifying this amplifying role in viral diffusion represents a significant advance in our understanding of social media information spread, moving beyond simple wide-reach effects to a more nuanced view of influencers' impact on content propagation.
	However, one possible explanation for these findings is that we have defined influencers via the hg index. Indeed, the results of the mixed-effects logistic regression (Table \ref{tab:regresults}) suggest that \texttt{sender\_follower\_counts} is not necessarily correlated with \texttt{sender\_influence}.
	%Specifically, while the coefficients for \texttt{sender\_follower\_counts} increase linearly, those for \texttt{sender\_influence} appear negative.
	This result indicates that a larger follower count does not automatically equate to greater influence (as operationalized by the hg index). Instead, these two measures seem to capture different facets of user impact, underscoring the importance of distinguishing between raw follower counts and influence metrics when evaluating an individual’s role in information diffusion.
	
	This research provides a bridge linking cultural evolution to a series of studies on online information diffusion from a computational social science perspective. Cultural evolution research has presented many insights into the adaptive aspects of human cognition through psychological experiments and mathematical simulations. Moreover, information diffusion research has presented quantitative analyses of the mechanisms of false information spread and echo chamber phenomena through the analysis of large-scale data from social network services. By integrating these two lines of research, our study confirmed that adaptive theories related to and hypotheses of cognition presented in cultural evolution theory can be observed even in social networks, a contemporary digital environment for human relationships. Furthermore, we demonstrated that prestige bias manifests in online communities, where information can be transmitted to others through easy \textit{sharing} functionalities.
	
	Our results emphasize the importance of influencers in marketing on social media \citep{Rogers2003,Kempe2003,Goel2012,Watts2007}. The fact that influencers’ influence is stronger for certain types of content (i.e., influencers’ impact on secondary spread varies depending on the content) can aid decision-making around influencer marketing strategies. These results implicitly suggest that influencers play a role in the spread of misinformation, which has become a significant social issue.
	
	Additionally, in recent years, researchers have attempted to propose solutions to the issue of false information from the perspectives of education and technology \citep{Toriumi2024} and behavioral science \citep{Lorenz-Spreen2020}.
	
	There is an argument that the spread of false information online is related to human behavior and cognition \cite{Acerbi2019}, and research has also shown that influencers can play an effective role in correcting false information \cite{Lim2022}. Given these considerations, the prestige bias highlighted in this study can be linked to one of the cognitive biases that influence information diffusion. From this perspective, this study may have potential applications in research on actual false information.
	
	Nevertheless, as this study demonstrates, the content of a post is also essential for reposting. Although this study did not distinguish between false information and other types of information, future research is needed to investigate how the correctness of information and cognitive biases interact.

	However, the content is still important. Therefore, we need to research interactions with content in future work.
	
	In conclusion, this study empirically demonstrated the existence and function of prestige bias in online communities and revealed that influencers have an impact on information diffusion through secondary spread.
	The concepts of primary and secondary spread proposed in this study, as well as the framework using virtual timelines and repost cascades, provide new perspectives for research on information diffusion in online communities. This research deepens our understanding of the influence of social media and provides insights into the complex interactions among user status, content sharing, and information propagation in online spaces. These findings can inform various applications, such as the development of more effective information diffusion strategies and measures against false information spread.
	%TC:ignore
	\section{Methods}
	\subsection{Data Collection and Preprocessing}
	
	We sampled Japanese-language reposts from Twitter (currently X) via the streaming API from October 1 to October 1 to 31, 2021. We chose Japanese posts as the objects of analysis for the following reasons:
	
	\begin{enumerate}
		\item The existence of a large Japanese-speaking community on Twitter
		\item A close correspondence between the Japanese language and national borders, enabling us to control for variations in cultural and social background
		\item A more limited and homogeneous sample than English-language posts
	\end{enumerate}
	
	Our one-month data collection period aligns with the timeframes used in previous research~\cite{Goel2012,Tsugawa2023}. We chose this timeframe for several reasons. First, our large-scale sampling method enabled us to gather a volume of data comparable to that of studies that utilized several years of information. Second, considering that the primary focus of this research is the relationship between influencers and instantaneous virality, a one-month period provides a sufficiently comprehensive snapshot to capture these dynamics.
	
	% 日本語のリツイートから取得されたことを追記
	To reconstruct follower–followee relationships as of October 2021, we used sampled users from users who had reposted Japanese posts, and we retrieved their following and follower lists every other day during September 2021.
	Notably, we were unable to capture private accounts, for which follower/following information is not publicly accessible. Our network reconstruction is thus limited to publicly visible connections and provides an approximation of the actual user links during the period; there may be some discrepancies with respect to actual follower‒followee relationships.
	
	Notably, our analysis focused solely on simple reposts on Twitter and excluded quote posts. This limitation is an important point to consider when interpreting the results.
	
	\subsection{Quantifying User Influence}
	\label{hg index}
	To quantify user influence, we adopted the hg index~\cite{Alonso2010}, which combines the h index~\cite{Hirsch2005} and g index~\cite{Egghe2006}. This metric was originally developed to evaluate scientific productivity, where citations are used as a measure of impact. In our application, we treat reposts of a user's posts as analogous to citations, allowing us to quantify a user's influence in social media contexts. This approach is able to balance consistent diffusion power (captured by the h index) and the scale of reposts (captured by the g index, similar to degree centrality).
	
	Although network centrality measures such as degree centrality and PageRank are commonly used to analyze the influence of source spreaders in social networks~\cite{Pei2014}, they may disproportionately emphasize users with a single viral post. \citet{Lü2016} demonstrated through a susceptible-infected-recovered (SIR) model that the h index, despite being strongly correlated with degree centrality and coreness, more appropriately evaluates node influence.
	However, the h index alone does not sufficiently account for the total number of reposts, which is important in social networks. Therefore, we adopted the hg index as a more robust measure that considers both the sustained influence (through the h index) and the engagement of individual posts (through the g index).
	
	The hg index was calculated via the following algorithm:
	
	\begin{enumerate}
		\item Sort all posts by a user in descending order of repost count
		\item Calculate the h index: the largest h where the h-th post has at least h reposts
		\item Calculate the g index: the largest g where the top g posts have at least $g^2$ total reposts
		\item hg index = $\sqrt{h \cdot g}$
	\end{enumerate}
	
	Importantly, the hg index mitigates the overestimation of influence on the basis solely of the total or maximum repost count, which can be skewed by a single viral post. Instead, it emphasizes users who achieve consistent engagement over time.
	
	On the basis of this metric, we classified users into six influence categories: very high (top 1\%), high (top 1--5\%), upper-mid (top 5--10\%), mid (top 10--30\%), lower-mid (top 30--50\%), and low (bottom 50\%). Note that these categories are exclusive of their upper thresholds, meaning, for example, that the \textit{high} category includes users above the 1\% threshold but below the 5\% threshold.
	
	\subsection{Construction and Analysis of Repost Cascades}
	Repost cascades (information cascades) model how information spreads across social networks. This concept, proposed by \citet{Bikhchandani1992} and further developed by \citet{Watts2002} and \citet{Kempe2003}, was applied in our study to capture repost chains on Twitter, following methodologies similar to those of \citet{Goel2012} and \citet{Vosoughi2018}.
	
	We constructed repost cascades using the following rules:
	
	\begin{enumerate}
		\item Set the original posts as the cascade root
		\item Select the temporally closest potential parent as the actual parent
		\item Exclude official accounts (identified by specific keywords in their account name, screen name, or profile)
	\end{enumerate}
	
	For a given repost, if there are multiple potential parent posts, we generally consider the repost with the most recent timestamp as the parent. This approach is based on our inference of how Twitter’s algorithm functioned at the time of the study.
	However, owing to the limitations of the X (formerly Twitter) API (which does not directly provide “who reposted from whom”) and the inability to track private accounts, it is not possible to precisely reconstruct the cascade. We therefore adopt this approximation method to infer the repost chain as closely as possible.
	Using this method, we constructed 4,882,985 repost cascades.
	
	By analyzing these cascades, we can understand how information propagates between users; we can determine the source of a user's repost through the cascade. Additionally, we can now calculate the depth and structural virality~\cite{Goel2016} of partial cascades formed by propagation from a specific reposted user.
	
	\subsection{Virtual Timeline}
	We constructed a virtual timeline from the sampled data and follower‒followee relationships, as illustrated in Figure \ref{fig:virtual_timeline_crp}(a). This approach is inspired by the method used by \citet{Vosoughi2018}. This virtual timeline simulates a single user's timeline, providing a realistic representation of how users encounter and interact with content when scrolling through social media feeds. Notably, in our study, this virtual timeline focuses exclusively on reposts.
	
	We arranged reposts by each user's followees in chronological order, applying the following conditions:
	\begin{enumerate}
		\item Include only users who made at least one repost during the period to remove dormant accounts
		\item Exclude official accounts (identified by specific keywords in their account name, screen name, or profile) from secondary spread cascades, as official users have different motivations for reposting. However, source posts from these accounts are not excluded from the analysis.
		\item Remove subsequent reposts by a user's followees for posts after the user reposts
	\end{enumerate}
	This method enables us to analyze what information users have been exposed to and how secondary spread occurs.
	
	The virtual timeline approach allows us to simulate users' exposure to content in a way that closely mimics real-world social media interactions. By focusing on reposts, we can more effectively track information diffusion through the network.
	
	\subsection{Cascading Repost Probability}
	To quantify the efficiency of information spread in secondary spread, we developed the concept of the CRP.
	
	First, we reconstructed each user’s timeline (virtual timeline) by considering the number of times they reposted and their follower relationships to estimate which reposts were visible to each user at a given time. Next, we traced whether a viewed repost leads to a subsequent repost (further diffusion) by modeling repost-based diffusion as an information cascade \citep{Goel2012, Vosoughi2018}. This approach allows us to count how many times a repost is viewed (i.e., appears on users’ timelines) and how many reposts are further shared. We then calculated, for each user, the total number of times their reposts were viewed (\textit{viewed reposts}), and the total number of times those \textit{viewed reposts} were further shared (\textit{reposted reposts}), obtaining a per-user CRP. Finally, we grouped users by their influence categories and aggregated the counts within each group, summing the total number of \textit{viewed reposts} and \textit{reposted reposts} contributed by all users in that group. Hence, we obtained a single CRP value for each group that captures how likely it is that reposts from that group are to be further shared, reflecting their overall effectiveness in secondary diffusion.
	
	Using the virtual timeline approach, we consider a post as reposted if it appears in a repost cascade, allowing us to calculate the probability of a user's secondary spread continuing, as shown in Figure \ref{fig:virtual_timeline_crp}(b).
	
	The CRP quantifies the likelihood of a repost being further shared, enabling us to evaluate the efficiency of repost diffusion by users with varying levels of influence.
	It thus provides a measure of a user's influence in the context of secondary spread, which can be formalized as follows.
	
	The CRP measures the likelihood of a repost being further shared and is defined as follows:
	\begin{equation}
		\text{CRP} = \frac{\text{Number of Reposted Reposts}}{\text{Number of Viewed Reposts}}
	\end{equation}
	
	As illustrated in Figure \ref{fig:virtual_timeline_crp}(c), for a single reposted user whose reposted content is viewed by three followers and further shared by two, the CRP would be $2/3 \approx 0.66$, indicating that 66\% of repost views led to further sharing.
	
	This approach allowed us to assess the extent to which users promote information spread, considering both the reach of their posts and the likelihood of those posts being further shared.
	
	\subsection{Mixed Effects Logistic Regression}
	We employed a mixed-effects logistic regression model to determine whether user influence is a significant driver of secondary reposts after accounting for both post- and user-level heterogeneity. In this analysis, we introduced random intercepts for the \texttt{source\_tweet\_id} (i.e., the original tweet) and the topic of the user profiles (\texttt{user\_topic}) learned by the biterm topic model\cite{Yan2013}. This approach helps control for unobserved differences in content-specific appeal and individual interests while allowing us to assess the influence of specific predictors on the likelihood of a repost.
	
	We used the \texttt{lme4} package\,\cite{Bates2015} in R to fit the model as follows:
	
	\begin{verbatim}
glmer(
  is_retweeted ~ followee_influence + repost_hour + sender_followers_count +
  user_followers_count + (1 | user_topic) + (1 | source_tweet_id),
  data = data,
  family = binomial(link = "logit"),
  nAGQ = 0,
  control = glmerControl(optimizer = "bobyqa", optCtrl = list(maxfun = 1e+05))
)
	\end{verbatim}
	The binary outcome variable, \texttt{is\_retweeted}, indicates whether the repost, which is shown to a user, triggered an additional repost by that user. As fixed effects, the model incorporates (1) the user's influence category (\texttt{sender\_influence}), coded via intercept (reference) coding with \textit{low} as the baseline; (2) the time of day the repost was shown (\texttt{repost\_hour}), coded with sum contrasts to capture diurnal patterns, which were referenced against Japan's National Statistics on media use\,\cite{MIAC2025}; (3) the sender's follower count (\texttt{sender\_followers\_count}); and (4) the viewer's follower count (\texttt{user\_followers\_count}). For both follower-count variables, we add 1 before applying the log transformation and then standardize the resulting values to facilitate model convergence and interpretation.
	
	To address the computational challenges arising from a vast dataset, we reduced the sample by collecting approximately 0.1\% of all positive (further reposted) cases and randomly drawing negative cases (displayed but not reposted) at twice the frequency of the positives. For comparability, negative instances were drawn only from tweets with a \texttt{source\_tweet\_id} present in the positive set and from timelines of users who appeared in the positive set. Because this sampling procedure artificially alters the original ratio of positive to negative outcomes, the model's intercept and coefficients cannot be interpreted as absolute probabilities of reposting. Nonetheless, they effectively capture \emph{relative} differences, such as contrasts among influence categories.
	
	Although a user-level random intercept such as \texttt{user\_id} would ordinarily capture individual differences, the data were too sparse at the single-user level---most users only reposted the same tweet once. Instead, we applied a random intercept for the topic of the user profiles to stabilize the estimation and account for higher-level differences reflecting user interests. The Supplementary Material S2 details this topic of user profiles. We also added a random intercept for \texttt{source\_tweet\_id} to control for postspecific characteristics influencing repost likelihood, such as content, emotional tone, or structural features. We used a binomial error distribution with a logit link function, specified \texttt{nAGQ} = 0, and employed the \texttt{bobyqa} optimizer to manage the computational load. % \ref{sup-user-profile}
	
	This model clarified the role of user influence in secondary spreading by integrating post and user random effects and controlling for key predictors such as the hour of the day, follower count, and influence category.
	
	\subsection{Max Depth and Structural Virality of First-Reposted Users}
	This analysis selected posts with more than 5,000 reposts to capture instances of substantial online diffusion. For each post, the first user who reshared it (the first-reposted user) was identified, and the corresponding repost cascade was isolated by tracing all subsequent shares. Users were categorized by influence level—ranging from low to very high—on the basis of criteria such as follower counts, previous engagement metrics, and other indicators of potential reach.
	
	To quantify the complexity and depth of the repost cascades, the first was structural virality, which, following the approach of \citet{Goel2016} and \citet{Vosoughi2018}, was computed as the average shortest path length among all pairs of users in a given cascade. Higher values of structural virality indicate multistep or chain-like diffusion, whereas lower values indicate broadcast-like patterns with fewer resharing steps. The second measure was the maximum depth, which was calculated by determining the longest path from the first reposted user to the final user in the chain, thus indicating how many layers of sharing emerged from the initial repost.
	
	These metrics were then compared across influence categories to assess whether highly influential users consistently spurred more extensive diffusion. By evaluating structural virality and maximum depth in the context of the initial user influence level, this study elucidated how early engagement by users with high influence can catalyze further resharing and produce cascading viral chains throughout the platform.
	
	\subsection{Ethical Considerations}
	As this study used only publicly available data and did not involve human subjects, it was exempt from ethics review according to the guidelines of the authors' affiliated institutions.
	
	\subsection{Limitations}
	\begin{itemize}
		\item \textbf{Language Constraints}\\
		The primary dataset used in this study consists of Japanese-language content.
		Although the English data we used cover a single day's worth of posts, this is insufficient for thorough experimentation.
		Furthermore, English data were not originally collected for this experiment.
		limiting its applicability to the current research goals.
		Consequently, comprehensive experiments and analyses involving English and other languages
		remain as future work.
		
		\item \textbf{Sampling Bias}\\
		The data analyzed in this study were collected via the X (formerly Twitter) API,
		meaning it represents a sampled subset rather than the platform’s entire dataset.
		Moreover, private accounts are excluded from data collection,
		which may introduce a certain degree of sampling bias.
		
		\item \textbf{Timeline Assumption}\\
		This study assumes that the timeline on X is chronological.
		Although we briefly discuss advertisements and algorithm-driven content reordering
		in the Supplementary Material S5, % \ref{sup-sensitivity-30mins}
		This consideration is not comprehensive.
	\end{itemize}
	
	\noindent
	These limitations should be considered when interpreting this study's findings and conclusions.
	Future work will aim to address these issues by expanding linguistic coverage,
	employ broader data collection methods,
	and more thoroughly examining the impact of algorithmic content ordering.
	
	\section{Acknowledgments}
	This work was supported by JSPS KAKENHI Grant Numbers JP22K18150 and JP23K28376.
	The English editing was provided by Editage (www.editage.jp), for which we are grateful.
	
	\section{Additional Information}
	\subsection{Competing interests}
	The authors declare that they have no competing interests.
	
	\section{Data and Code availability}
	All datasets and codes used in this work are available at \url{https://github.com/asahi-research/secondary_spreads_of_influencers}.

	\bibliographystyle{unsrtnat}
	\bibliography{reference.bib}

\begin{thebibliography}{36}
\providecommand{\natexlab}[1]{#1}
\providecommand{\url}[1]{\texttt{#1}}
\expandafter\ifx\csname urlstyle\endcsname\relax
  \providecommand{\doi}[1]{doi: #1}\else
  \providecommand{\doi}{doi: \begingroup \urlstyle{rm}\Url}\fi

\bibitem[Pei et~al.(2014)Pei, Muchnik, Andrade~Jr., Zheng, and Makse]{Pei2014}
Sen Pei, Lev Muchnik, Jos{\'e}~S. Andrade~Jr., Zhiming Zheng, and Hern{\'a}n~A. Makse.
\newblock {Searching for superspreaders of information in real-world social media}.
\newblock \emph{Scientific Reports}, 4\penalty0 (1):\penalty0 5547, 2014.

\bibitem[Kempe et~al.(2003)Kempe, Kleinberg, and Tardos]{Kempe2003}
David Kempe, Jon Kleinberg, and \'{E}va Tardos.
\newblock {Maximizing the spread of influence through a social network}.
\newblock In \emph{Proceedings of the Ninth ACM SIGKDD International Conference on Knowledge Discovery and Data Mining}, pages 137--146, 2003.

\bibitem[Costello and Yesiloglu(2020)]{Yesiloglu2020}
Joyce Costello and Sevil Yesiloglu, editors.
\newblock \emph{{Influencer Marketing Building Brand Communities and Engagement}}.
\newblock Routledge, 2020.

\bibitem[Zietek(2016)]{Zietek2016}
Nathalie Zietek.
\newblock Influencer marketing : the characteristics and components of fashion influencer marketing.
\newblock Master's thesis, {University of Borås, Faculty of Textiles, Engineering and Business}, 2016.

\bibitem[Watts and Peretti(2007)]{Watts2007}
Duncan~J. Watts and Jonah Peretti.
\newblock {Viral marketing for the real world}.
\newblock \emph{Harvard Business Review}, 85\penalty0 (5), 2007.

\bibitem[Murakami(2021)]{Murakami2021}
Yuri Murakami.
\newblock {Artist, retweeters ordered to pay journalist over ‘fake rape’ posts}, 2021.
\newblock URL \url{https://www.asahi.com/ajw/articles/14492778}.
\newblock Accessed on September 26, 2024.

\bibitem[Lim et~al.(2022)Lim, Toriumi, and Yoshida]{Lim2022}
Dongwoo Lim, Fujio Toriumi, and Mitsuo Yoshida.
\newblock {Do you trust experts on Twitter? Successful correction of COVID-19-related misinformation}.
\newblock In \emph{IEEE/WIC/ACM International Conference on Web Intelligence and Intelligent Agent Technology}, pages 518--523, 2022.

\bibitem[Morone and Makse(2015)]{Morone2015}
Flaviano Morone and Hern{\'a}n~A. Makse.
\newblock {Influence maximization in complex networks through optimal percolation}.
\newblock \emph{Nature}, 524\penalty0 (7563):\penalty0 65--68, 2015.

\bibitem[Tsugawa and Watabe(2023)]{Tsugawa2023}
Sho Tsugawa and Kohei Watabe.
\newblock {Identifying Influential Brokers on Social Media from Social Network Structure}.
\newblock \emph{Proceedings of the International AAAI Conference on Web and Social Media}, 17\penalty0 (1):\penalty0 842--853, 2023.

\bibitem[Acerbi(2022)]{Acerbi2022}
Alberto Acerbi.
\newblock {From Storytelling to Facebook}.
\newblock \emph{Human Nature}, 33\penalty0 (2):\penalty0 132--144, 2022.

\bibitem[Burt(2000)]{Burt2000}
Ronald~S. Burt.
\newblock {The Network Structure Of Social Capital}.
\newblock \emph{Research in Organizational Behavior}, 22:\penalty0 345--423, 2000.

\bibitem[Theo~Araujo and Vliegenthart(2017)]{TheoAraujo2017}
Peter~Neijens Theo~Araujo and Rens Vliegenthart.
\newblock {Getting the word out on Twitter: the role of influentials, information brokers and strong ties in building word-of-mouth for brands}.
\newblock \emph{International Journal of Advertising}, 36\penalty0 (3):\penalty0 496--513, 2017.

\bibitem[Henrich and Gil-White(2001)]{Henrich2001}
Joseph Henrich and Francisco~J. Gil-White.
\newblock {The evolution of prestige: freely conferred deference as a mechanism for enhancing the benefits of cultural transmission}.
\newblock \emph{Evolution and Human Behavior}, 22\penalty0 (3):\penalty0 165--196, 2001.

\bibitem[Jim{\'e}nez and Mesoudi(2019)]{Jimenez2019}
{\'A}ngel~V. Jim{\'e}nez and Alex Mesoudi.
\newblock {Prestige-biased social learning: current evidence and outstanding questions}.
\newblock \emph{Palgrave Communications}, 5\penalty0 (1):\penalty0 20, 2019.

\bibitem[Henrich and McElreath(2003)]{Henrich2003}
Joseph Henrich and Richard McElreath.
\newblock {The evolution of cultural evolution}.
\newblock \emph{Evolutionary Anthropology: Issues, News, and Reviews}, 12\penalty0 (3):\penalty0 123--135, 2003.

\bibitem[Katz et~al.(1955)Katz, Lazarsfeld, and Roper]{Katz1955}
Elihu Katz, Paul~F. Lazarsfeld, and Elmo Roper.
\newblock \emph{{Personal Influence: The Part Played by People in the Flow of Mass Communications}}.
\newblock The Free Press, 1955.

\bibitem[Rogers(2003)]{Rogers2003}
Everett~M. Rogers.
\newblock \emph{{Diffusion of Innovations, 5th Edition}}.
\newblock Free Press, 2003.

\bibitem[Goel et~al.(2012)Goel, Watts, and Goldstein]{Goel2012}
Sharad Goel, Duncan~J. Watts, and Daniel~G. Goldstein.
\newblock {The structure of online diffusion networks}.
\newblock In \emph{Proceedings of the 13th ACM Conference on Electronic Commerce}, pages 623--638, 2012.

\bibitem[Bikhchandani et~al.(1992)Bikhchandani, Hirshleifer, and Welch]{Bikhchandani1992}
Sushil Bikhchandani, David Hirshleifer, and Ivo Welch.
\newblock {A theory of fads, fashion, custom, and cultural change as informational cascades}.
\newblock \emph{Journal of Political Economy}, 100\penalty0 (5):\penalty0 992--1026, 1992.

\bibitem[Watts(2002)]{Watts2002}
Duncan~J. Watts.
\newblock {A simple model of global cascades on random networks}.
\newblock \emph{Proceedings of the National Academy of Sciences}, 99\penalty0 (9):\penalty0 5766--5771, 2002.

\bibitem[Vosoughi et~al.(2018)Vosoughi, Roy, and Aral]{Vosoughi2018}
Soroush Vosoughi, Deb Roy, and Sinan Aral.
\newblock {The spread of true and false news online}.
\newblock \emph{Science}, 359\penalty0 (6380):\penalty0 1146--1151, 2018.

\bibitem[Alonso et~al.(2010)Alonso, Cabrerizo, Herrera-Viedma, and Herrera]{Alonso2010}
Sergio Alonso, Fancisco.~Javier Cabrerizo, Enrique Herrera-Viedma, and Francisco Herrera.
\newblock {hg-index: a new index to characterize the scientific output of researchers based on the h- and g-indices}.
\newblock \emph{Scientometrics}, 82\penalty0 (2):\penalty0 391--400, 2010.

\bibitem[Hirsch(2005)]{Hirsch2005}
Jorge~E Hirsch.
\newblock {An index to quantify an individual's scientific research output}.
\newblock \emph{Proceedings of the National Academy of Sciences}, 102\penalty0 (46):\penalty0 16569--16572, 2005.

\bibitem[Yamaguchi et~al.(2015)Yamaguchi, Yoshida, Faloutsos, and Kitagawa]{Yamaguchi2015}
Yuto Yamaguchi, Mitsuo Yoshida, Christos Faloutsos, and Hiroyuki Kitagawa.
\newblock Patterns in interactive tagging networks.
\newblock \emph{Proceedings of the International AAAI Conference on Web and Social Media}, 9\penalty0 (1):\penalty0 513–522, August 2015.
\newblock ISSN 2162-3449.
\newblock \doi{10.1609/icwsm.v9i1.14616}.
\newblock URL \url{http://dx.doi.org/10.1609/icwsm.v9i1.14616}.

\bibitem[Yoshida and Yamaguchi(2015)]{Yoshida2015}
Mitsuo Yoshida and Yuto Yamaguchi.
\newblock Interactive tagging networks (following/followers and tags on 1 million twitter users), 2015.
\newblock URL \url{https://zenodo.org/record/16267}.

\bibitem[Acerbi and Tehrani(2018)]{Acerbi2018}
Alberto Acerbi and Jamshid~J. Tehrani.
\newblock {Did Einstein Really Say that? Testing Content Versus Context in the Cultural Selection of Quotations}.
\newblock \emph{Journal of Cognition and Culture}, 18\penalty0 (3-4):\penalty0 293--311, 2018.

\bibitem[Brand et~al.(2021)Brand, Mesoudi, and Morgan]{Brand2021}
Charlotte~O. Brand, Alex Mesoudi, and Thomas J.~H. Morgan.
\newblock {Trusting the experts: The domain-specificity of prestige-biased social learning}.
\newblock \emph{PLOS ONE}, 16\penalty0 (8):\penalty0 1--15, 2021.

\bibitem[Goel et~al.(2016)Goel, Anderson, Hofman, and Watts]{Goel2016}
Sharad Goel, Ashton Anderson, Jake Hofman, and Duncan~J. Watts.
\newblock {The structural virality of online diffusion}.
\newblock \emph{Manage. Sci.}, 62\penalty0 (1):\penalty0 180--196, 2016.

\bibitem[Toriumi and Yamamoto(2024)]{Toriumi2024}
Fujio Toriumi and Tatsuhiko Yamamoto.
\newblock {Informational Health --Toward the Reduction of Risks in the Information Space}, 2024.
\newblock URL \url{https://arxiv.org/abs/2407.14634}.

\bibitem[Lorenz-Spreen et~al.(2020)Lorenz-Spreen, Lewandowsky, Sunstein, and Hertwig]{Lorenz-Spreen2020}
Philipp Lorenz-Spreen, Stephan Lewandowsky, Cass~R. Sunstein, and Ralph Hertwig.
\newblock {How behavioural sciences can promote truth, autonomy and democratic discourse online}.
\newblock \emph{Nat. Hum. Behav.}, 4\penalty0 (11):\penalty0 1102--1109, 2020.

\bibitem[Acerbi(2019)]{Acerbi2019}
Alberto Acerbi.
\newblock Cognitive attraction and online misinformation.
\newblock \emph{Palgrave Communications}, 5\penalty0 (1), February 2019.
\newblock ISSN 2055-1045.
\newblock \doi{10.1057/s41599-019-0224-y}.
\newblock URL \url{http://dx.doi.org/10.1057/s41599-019-0224-y}.

\bibitem[Egghe(2006)]{Egghe2006}
Leo Egghe.
\newblock {Theory and practise of the g-index}.
\newblock \emph{Scientometrics}, 69\penalty0 (1):\penalty0 131--152, 2006.

\bibitem[L{\"u} et~al.(2016)L{\"u}, Zhou, Zhang, and Stanley]{Lü2016}
Linyuan L{\"u}, Tao Zhou, Qian-Ming Zhang, and H.~Eugene Stanley.
\newblock {The H-index of a network node and its relation to degree and coreness}.
\newblock \emph{Nature Communications}, 7\penalty0 (1):\penalty0 10168, 2016.

\bibitem[Yan et~al.(2013)Yan, Guo, Lan, and Cheng]{Yan2013}
Xiaohui Yan, Jiafeng Guo, Yanyan Lan, and Xueqi Cheng.
\newblock A biterm topic model for short texts.
\newblock In \emph{Proceedings of the 22nd international conference on World Wide Web}, WWW ’13, page 1445–1456. ACM, May 2013.
\newblock \doi{10.1145/2488388.2488514}.
\newblock URL \url{http://dx.doi.org/10.1145/2488388.2488514}.

\bibitem[Bates et~al.(2015)Bates, M\"{a}chler, Bolker, and Walker]{Bates2015}
Douglas Bates, Martin M\"{a}chler, Ben Bolker, and Steve Walker.
\newblock Fitting linear mixed-effects models usinglme4.
\newblock \emph{Journal of Statistical Software}, 67\penalty0 (1), 2015.
\newblock ISSN 1548-7660.
\newblock \doi{10.18637/jss.v067.i01}.
\newblock URL \url{http://dx.doi.org/10.18637/jss.v067.i01}.

\bibitem[for Information et~al.(2023)for Information, Communications~Policy, and (Japan)]{MIAC2025}
Institute for Information, Ministry of Internal~Affairs Communications~Policy, and Communications (Japan).
\newblock Survey report (summary version) on time spent on information and communications media and on information behavior (fiscal year 2022) (japanese), 2023.
\newblock URL \url{https://www.soumu.go.jp/main_content/000887659.pdf}.
\newblock Accessed on February 10, 2025.

\end{thebibliography}


\begin{thebibliography}{4}
\providecommand{\natexlab}[1]{#1}
\providecommand{\url}[1]{\texttt{#1}}
\expandafter\ifx\csname urlstyle\endcsname\relax
  \providecommand{\doi}[1]{doi: #1}\else
  \providecommand{\doi}{doi: \begingroup \urlstyle{rm}\Url}\fi

\bibitem[Bates et~al.(2015)Bates, M\"{a}chler, Bolker, and Walker]{Bates2015}
Douglas Bates, Martin M\"{a}chler, Ben Bolker, and Steve Walker.
\newblock Fitting linear mixed-effects models usinglme4.
\newblock \emph{Journal of Statistical Software}, 67\penalty0 (1), 2015.
\newblock ISSN 1548-7660.
\newblock \doi{10.18637/jss.v067.i01}.
\newblock URL \url{http://dx.doi.org/10.18637/jss.v067.i01}.

\bibitem[Kudo(2005)]{Kudo2005}
Taku Kudo.
\newblock Mecab : Yet another part-of-speech and morphological analyzer.
\newblock \emph{http://mecab.sourceforge.net/}, 2005.
\newblock URL \url{https://cir.nii.ac.jp/crid/1572543025344508032}.

\bibitem[Yamaguchi et~al.(2015)Yamaguchi, Yoshida, Faloutsos, and Kitagawa]{Yamaguchi2015}
Yuto Yamaguchi, Mitsuo Yoshida, Christos Faloutsos, and Hiroyuki Kitagawa.
\newblock Patterns in interactive tagging networks.
\newblock \emph{Proceedings of the International AAAI Conference on Web and Social Media}, 9\penalty0 (1):\penalty0 513–522, August 2015.
\newblock ISSN 2162-3449.
\newblock \doi{10.1609/icwsm.v9i1.14616}.
\newblock URL \url{http://dx.doi.org/10.1609/icwsm.v9i1.14616}.

\bibitem[Yan et~al.(2013)Yan, Guo, Lan, and Cheng]{Yan2013}
Xiaohui Yan, Jiafeng Guo, Yanyan Lan, and Xueqi Cheng.
\newblock A biterm topic model for short texts.
\newblock In \emph{Proceedings of the 22nd international conference on World Wide Web}, WWW ’13, page 1445–1456. ACM, May 2013.
\newblock \doi{10.1145/2488388.2488514}.
\newblock URL \url{http://dx.doi.org/10.1145/2488388.2488514}.

\end{thebibliography}
	
	\section{Author contributions}
	T.N. implemented the code and analyzed the data.
	Y.N., M.Y., and H.T. supervised the research.
	Y.N. guided the hypothesis formulation and interpretation of experimental results, M.Y. guided the design of analytical models and computational frameworks, and H.T. guided the setup of the computational experimental environment.
	M.Y. collected and managed the research data.
	H.T. was responsible for setting up the computational environment.
	T.N. and Y.N. wrote the main manuscript.
	All authors reviewed the analytical methodology, discussed the results, and contributed to the final manuscript.
	%TC:endignore
\end{document}

\maketitle
\makeatletter
% \linenumbers

\renewcommand \thesection{S\@arabic\c@section}
\renewcommand\thetable{S\@arabic\c@table}
\renewcommand \thefigure{S\@arabic\c@figure}
\makeatother

\section{Detail of English Dataset}
\label{sup-english-data}
In addition to our main dataset analyzed in this study, we also examined an English-language dataset to check the robustness of our findings. Specifically, we used a set of one million users (sampled in December 2014 from the global timeline) provided by \citet{Yamaguchi2015} and collected all reposts over 24 hours on January 1, 2015.
Note that because these users were sampled from the global timeline, they may not all be English speakers.

As in our main analysis, we applied the same hg index-based approach to measure user influence. However, because this dataset covers only a single day, the resulting hg index values are less fine-grained than our main dataset, which spans 31 days of Japanese posts. Consequently, we classified users into four categories (low, mid, high, and very high) rather than six.

Furthermore, owing to the limited temporal coverage (24 hours), we analyzed the cascading repost probability (CRP) in shorter time windows (30 minutes, 1 hour, and 3 hours). Although the observation period is more constrained than our main analysis, we observe similar trends: users in the very high influence group consistently achieve higher CRP. We also note that some typically influential users may be underrepresented or misclassified due to the brief window. Therefore, we should interpret these results as a supplemental check rather than a comprehensive analysis of long-term user behavior and diffusion dynamics.

\section{Topics of User Profiles}
\label{sup-user-profile}
User profiles used for the regression analysis were tokenized using the Japanese tokenizer Mecab~\cite{Kudo2005}, splitting the text into word-level tokens. As preprocessing, we removed stop words, function words, URLs, and mentions. After this, we applied \textbf{Biterm Topic Modeling (BTM)}\cite{Yan2013}, which is particularly well suited for short texts. For simplicity, we set the number of topics to 10. Table~\ref{tab:user_topics} provides an overview of these 10 topics, including example Japanese keywords and their approximate English translations for reference.

\begin{CJK*}{UTF8}{ipxm}
\begin{table}[htbp]
\centering
\caption{Overview of 10 Topics of User's Profiles with Example Keywords (Japanese). English keywords in parentheses are provided for reference.}
\label{tab:user_topics}
\begin{tabularx}{\textwidth}{c p{4cm} X}
\toprule
\textbf{Topic} & \textbf{Label} & \textbf{Example Keywords (JP / EN)} \\
\midrule
1 & Fandoms of Male Artists
  & ハート (heart), 応援 (support), ファン (fan), SMAP (artist name), 嵐 (Arashi / artist name) \\
\midrule
2 & Raffles and Others
  & 懸賞 (raffle), 当選 (winner), 旅行 (travel), 猫 (cat), コスメ (cosmetics) \\
\midrule
3 & English Phrases
  & I, AND, of, TO, IS \\
\midrule
4 & Gaming and Anime Fandoms 
  & ウマ娘 (Uma Musume / game title), FGO (game title), アイコン (icon), 描く (draw), ポケモン (Pokemon / game title) \\
\midrule
5 & Political Views and Opinions 
  & 日本 (Japan), 反対 (oppose), 政治 (politics), 自分 (self), コロナ (COVID-19) \\
\midrule
6 & Fandoms For Female Content
  & 腐る (decay), FGO (Game) , 夢 (dream), ツイステッドワンダーランド (Twisted Wonderland / game title), 腐女子 (fujoshi) \\
\midrule
7 & Bussines Use
  & DM (direct mail), 情報 (information), 依頼 (request), お仕事 (job), イベント (event) \\
\midrule
8 & Sports and Entertainment Fandoms 
  & 野球 (baseball), サッカー (soccer), 音楽 (music), ライブ (live), 乃木坂46 (Nogizaka46 / artist name) \\
\midrule
9 & Drawing Community
  & 描く (draw), 絵 (picture), アイコン (icon), ツイート (tweet), イラスト (illustration) \\
\midrule
10 & Miscellaneous
  & 趣味 (hobby), 音楽 (music), 映画 (movie), 写真 (photography), 漫画 (manga) \\
\bottomrule
\end{tabularx}
\end{table}
\end{CJK*}

\section{Topic Analysis on Random Effects}
\label{sup-content-analysis}
This section describes the random effects of the mixed-effects model in subsection 2.4, mainly focusing on posts.
Additionally, we provide an analysis of the topics of these posts.

In our main analysis, we fitted a mixed‐effects model using the \texttt{lme4}\cite{Bates2015} package in R, specifying random intercepts for both \texttt{source\_tweet\_id} and \texttt{user\_topic}.
Table~\ref{tab:randomeffects} summarizes the random effects that Table 2 in the main analysis.
The standard deviation for \texttt{source\_tweet\_id} (0.7733) is higher than that for \texttt{user\_topic} (0.4321).
This indicates that each tweet captured by \texttt{source\_tweet\_id} contributes more to whether a post reposts than differences among user topics.
Notably,  the standard deviation of \texttt{source\_tweet\_id} is larger than the coefficients of \texttt{sender\_influence} for \textit{very high} users.

\begin{table}[htbp]
\centering
\caption{Random effects from the mixed‐effects model.}
\label{tab:randomeffects}
\begin{tabular}{l l r r}
\toprule
\textbf{Groups} & \textbf{Name} & \textbf{Variance} & \textbf{Std.Dev.}\\
\midrule
\texttt{source\_tweet\_id} & (Intercept) & 0.5980 & 0.7733 \\
\texttt{user\_topic}       & (Intercept) & 0.1868 & 0.4321 \\
\bottomrule
\end{tabular}
\end{table}

To explore whether topical differences might explain the sizeable post‐level variation, we applied the same model (BTM) for the topics of user profiles to the dataset of source tweets. We tested 10, 15, and 30 topics, and the 10‐topic model yielded the highest coherence. Table~\ref{tab:post_topics} provides an overview of these 10 topics, including example Japanese keywords and approximate English translations for reference.

\begin{figure}[htbp]
    \centering
    \includegraphics[width=0.7\textwidth]{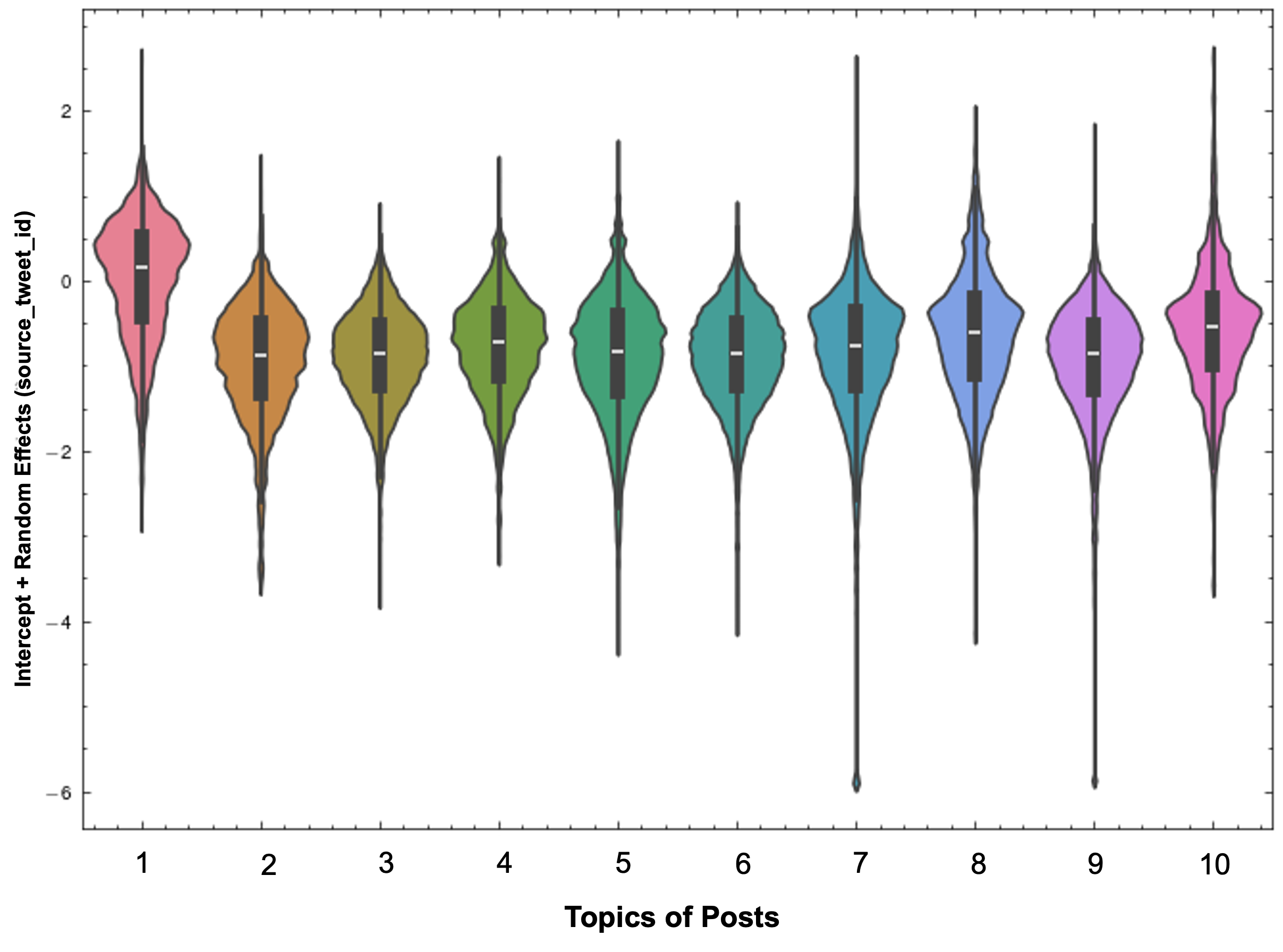}
    \caption{Sample visualization of intercept + random effects for each topic.}
    \label{fig:random_effect_plot}
\end{figure}

\begin{CJK*}{UTF8}{ipxm}
\begin{table}[htbp]
\centering
\caption{Overview of 10 Topics of Posts with Example Keywords (Japanese). English keywords in parentheses are provided for reference.}
\label{tab:post_topics}
\begin{tabularx}{\textwidth}{c p{4cm} X}
\toprule
\textbf{Topic} & \textbf{Label} & \textbf{Example Keywords (JP / EN)} \\
\midrule
1 & Promotional Campaigns \& Giveaways 
  & プレゼント (present), フォロー (follow), キャンペーン (campaign) \\
\midrule
2 & Miscellaneous Daily Life 
  & 猫 (cat), 犬 (dog), 食べる (eat), 地震 (earthquake), 水 (water) \\
\midrule
3 & Political / Election-Related 
  & 選挙 (election), 投票 (vote), 政党 (political party), 政権交代 (change of government) \\
\midrule
4 & Sports Events (Baseball, Soccer, etc.)
  & 選手 (player), 試合 (match), 優勝 (victory), チーム (team) \\
\midrule
5 & Creative Works, Illustrations, Gaming
  & イラスト (illustration), ゲーム (game), デザイン (design), 描く (draw) \\
\midrule
6 & COVID-19 and Government Policy
  & ワクチン (vaccine), 接種 (inoculation), コロナ (COVID-19), 政府 (government) \\
\midrule
7 & Positive Emotions and Festive Themes
  & 笑顔 (smile), ハロウィン (Halloween), 嬉しい (happy), 誕生日 (birthday) \\
\midrule
8 & Media, Broadcasting, and Streaming
  & 配信 (streaming), 放送 (broadcast), 映画 (movie), 動画 (video) \\
\midrule
9 & General Opinions and Everyday Reflections
  & 思う (think), 言う (say), 自分 (myself), 見る (see) \\
\midrule
10 & Event Announcements and Ticketing
   & 開催 (holding an event), チケット (ticket), 予約 (reservation), 会場 (venue) \\
\bottomrule
\end{tabularx}
\end{table}
\end{CJK*}

Figure \ref{fig:random_effect_plot} shows the model’s intercept plus its random effect for each \texttt{source\_tweet\_id}, illustrating whether these distributions differ substantially by topic.
In short, no topic stands out as having an especially high or low average repost likelihood. This suggests that, while the unique content of each post strongly influences its chances of being reposted, it is not simply a matter of belonging to one of the 10 broad topics. Consequently, factors such as emotional valence, writing style, timing, or visual elements may play larger roles in driving repost behavior. These considerations remain open areas for future investigation.

% 多重共線性がないことを書いておくべき？

\section{Analysis of Influencers' Reposting Behavior}
\label{sup-reposting-behavior}

\begin{figure}[htbp]
    \centering
    \includegraphics[width=\textwidth]{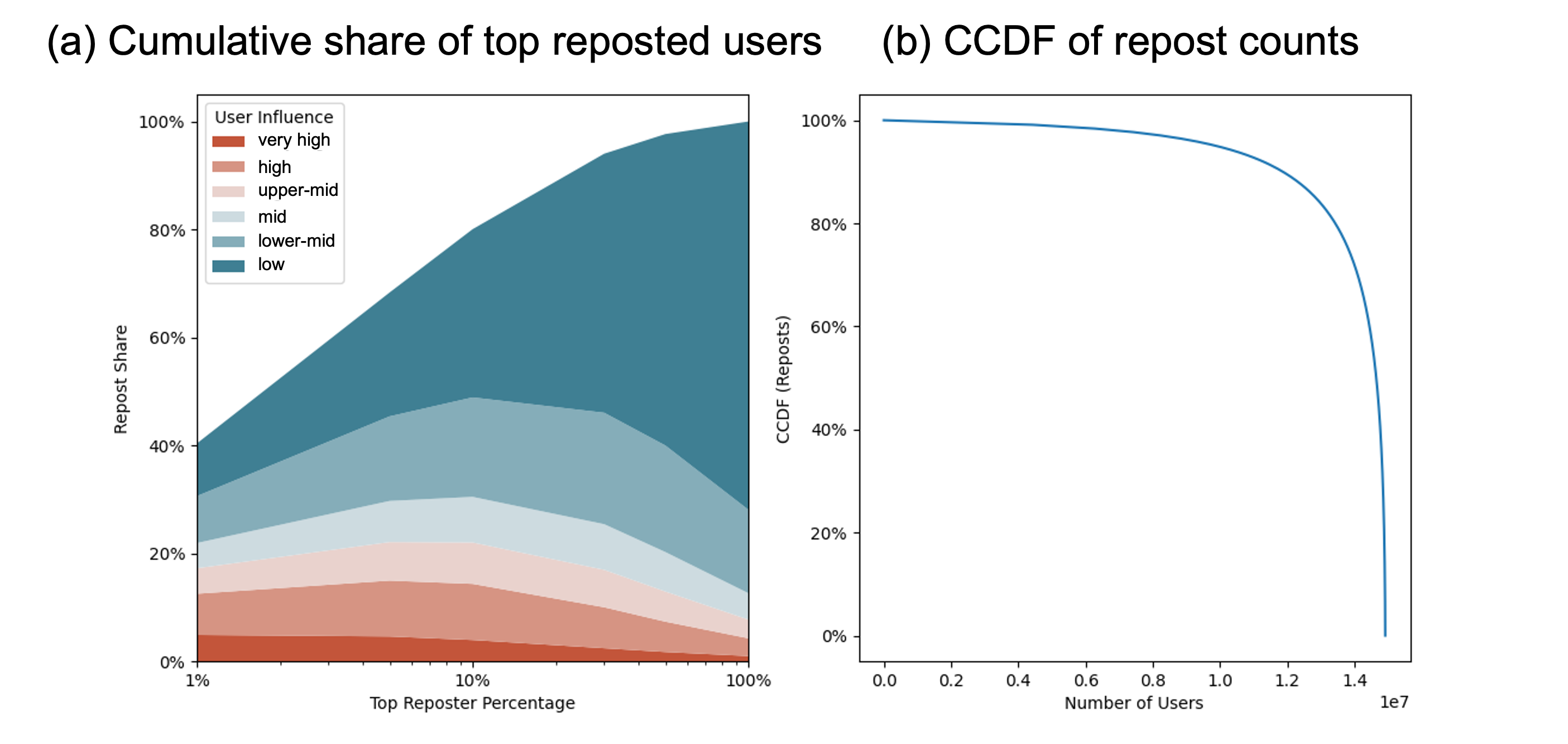}
    \caption{Analysis of users' reposting behavior. (a) Cumulative share of top reposted users. (b) CCDF of repost counts.}
    \label{fig:repost_behavior}
\end{figure}

To further investigate the cause of influencers' high share in the secondary spread, we analyzed users' reposting behavior by aggregating the number of reposts for each user and visualizing the proportion of reposts accounted for by top-reposted users (Figure \ref{fig:repost_behavior}).

Figure \ref{fig:repost_behavior}(a) reveals that the top 1\% of reposted users account for 30\% of all reposts, whereas the top 20\% account for 80\%. Importantly, these top-reposted users comprise users with various levels of influence. The Complementary Cumulative Distribution Function (CCDF) in Figure \ref{fig:repost_behavior}(b) demonstrates that while the majority of users make very few reposts, a small number of users make a disproportionately large number of reposts. 

These findings suggest that the high share of influencers in the secondary spread is not merely due to their higher frequency of reposting. This result indicates that the influence category significantly affects the information distribution efficiency in the secondary spread. Information reposted by influencers tends to reach a wider audience and is more likely to be reposted than information shared by other user groups, which is consistent with prestige bias. 

Additionally, the results show a substantial skew in the number of reposts among users.
Therefore, a little of users spread the most of reposts,
indicating potential bottlenecks in information diffusion and the risks of excessively spreading certain information.
These results and our main results demonstrate that influencers' high share in the secondary spread is attributed to the qualitative aspects of their influence rather than simply a higher volume of reposts.
This analysis highlights the issue of skewed information diffusion as a new area for future research.

\section{Sensitivity Analysis Within 30 Minutes of Posting}
\label{sup-sensitivity-30mins}

Although our primary analyses used a virtual timeline constructed in a near-chronological manner, we acknowledge that X (formerly Twitter) provides algorithmic recommendations such as trending topics, which can reorder or highlight posts in ways that deviate from strict chronological sequence.
To reduce these algorithmic effects, we limit our analysis to the first 30 minutes after a post’s publication.
Trending or recommended content typically relies on accumulating engagement signals over a longer timeframe.

\begin{figure}[h]
\centering \includegraphics[width=0.45\textwidth]{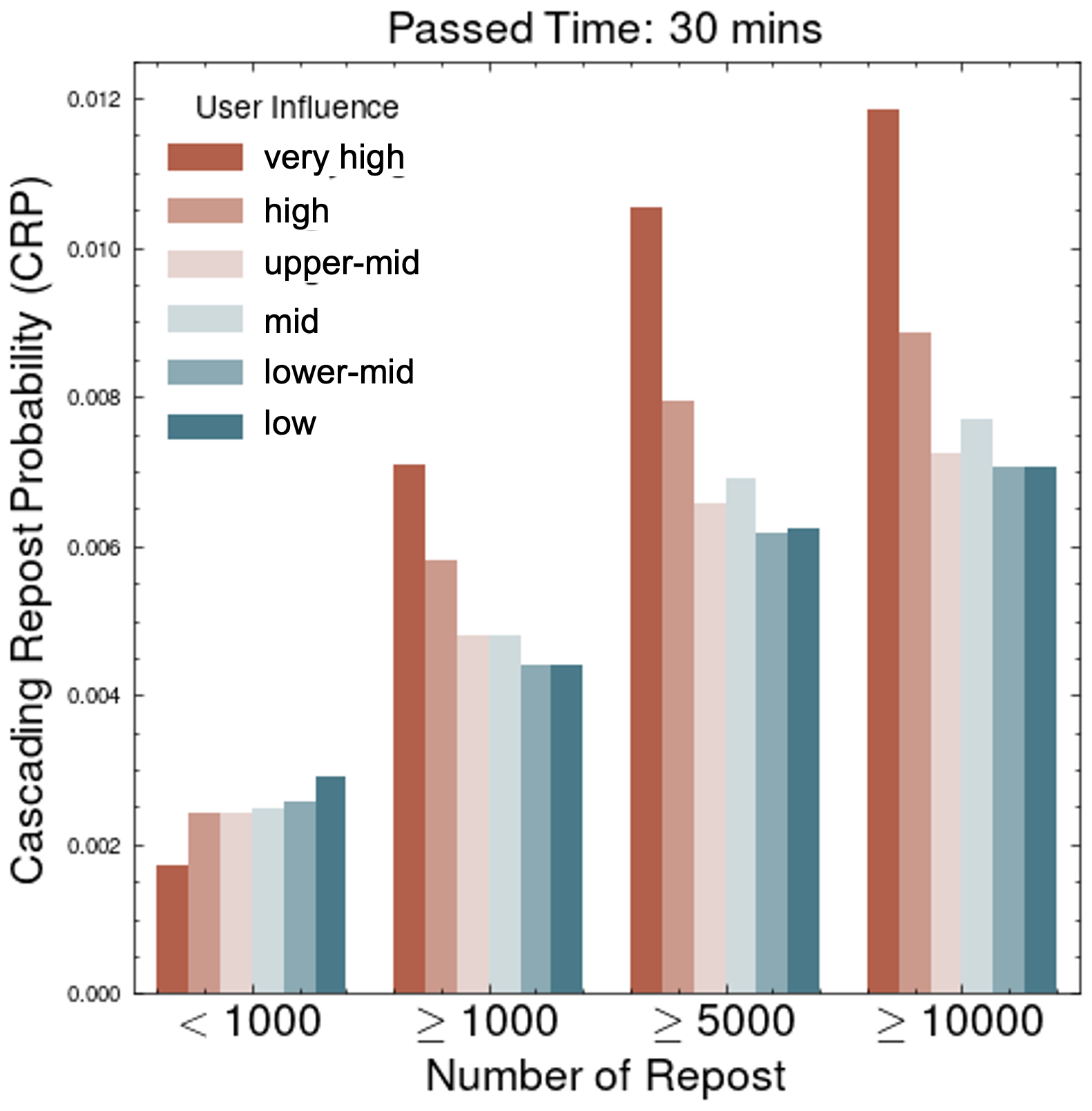}
\caption{Cascading Repost Probability (CRP) within 30 minutes of the original post, broken down by user influence categories and post popularity thresholds.}
\label{fig:30min_analysis}
\end{figure}

Figure~\ref{fig:30min_analysis} shows that even in this shorter timeframe—where the influence of algorithmic ordering may be relatively limited—the observed trends in repost behavior remain consistent with our main findings. Specifically, higher influence users continue to exhibit relatively higher CRP values for popular posts (e.g., 
$\geq1000$ reposts), suggesting that the prestige bias effect is robust under these near-chronological conditions.

We note, however, that some algorithmic prioritization could still occur within these initial 30 minutes. Nevertheless, the consistency of these results supports the view that our near-chronological virtual timeline reasonably approximates user exposure patterns, capturing meaningful insights about information diffusion and user influence in the early stages of reposting.

\bibliographystyle{plainnat}
\bibliography{reference.bib}